\documentclass[twocolumn,superscriptaddress,showpacs,preprintnumbers,amsmath,amssymb]{revtex4}
\usepackage{comment}
\usepackage{ulem}

\usepackage{amsthm}
\usepackage{mathtools}
\usepackage{color}
\usepackage{graphicx}
\usepackage[dvipdfmx]{hyperref}
\usepackage{bm}
\usepackage{mathrsfs}
\usepackage{amsfonts}

\begin{document}

\title{Demon driven by geometric phase}

\author{Ryosuke Yoshii}\email[e-mail address]{ryoshii@rs.socu.ac.jp}
\affiliation{Center for Liberal Arts and Sciences, Sanyo-Onoda City University,
Yamaguchi 756-0884, Japan}

\author{Hisao Hayakawa}\email{hisao@yukawa.kyoto-u.ac.jp}
\affiliation{Center for Gravitational Physics and Quantum Information, Yukawa Institute for Theoretical Physics, Kyoto University, Kitashirakawa-oiwake cho, Sakyo-ku, Kyoto 606-8502, Japan}

\date{\today}

\begin{abstract} 
We theoretically study the entropy production and the work extracted from a system connected to two reservoirs by periodic modulations of the electrochemical potentials of the reservoirs and the parameter of a system Hamiltonian under isothermal conditions. 
We find that the modulation of the parameters can drive a geometric state, which is away from a nonequilibrium steady state. 
Using this property, we construct a demon in which the entropy production during the first one-cycle is negative such that we can extract the work 
if we start from the nonequilibrium steady state without parameter modulations. 
We use the Anderson model to implement the demon in a realistic situation. 

\end{abstract}
\maketitle

\section{Introduction}

The second law of thermodynamics is one of the most fundamental laws of physics, 
specifying the upper limit of the available work that can be extracted from reservoirs.
Maxwell proposed an ideal setup to violate the second law, 
in which a demon quickly opens and closes the gate, allowing only fast-moving molecules to pass through in one direction~\cite{Maxwell}. 
This leads to a decrease in entropy without any work being done and violates the second law of thermodynamics.
A simplified version of Maxwell's demon was proposed by Szilard~\cite{Szilard}. 
It was shown that Maxwell's demon can be used to implement the Szilard engine, 
while the original Maxwell's demon makes the temperature gradient act like a refrigerator. 
Since then, the demon problem has attracted much attention from physicists~\cite{Brillouin51a,Brillouin51b,Brillouin56,Jaynes57,Landauer61,Bennett73}.
Modern experiments use photonic devices to realise Maxwell's demon \cite{Koski14,Vidrighin16,Shu17}.
A comprehensive historical review of Maxwell's demon is given in Ref.\,\cite{Rex17}.

Since Maxwell's original idea was based on the measurement of molecules, 
it is natural to combine the physical science with information thermodynamics to realise Maxwell's demon~\cite{Brillouin51a,Brillouin51b,Brillouin56,Jaynes57,Landauer61,Bennett73,Koski14,Touchette00,Sagawa10,Sagawa12,Leigh07,Toyabe10,Parrondo15}.
However, the cost of implementing the informational Maxwell's demon is high, although the theoretical formulation ignores this cost. 
Instead, we propose a geometric demon using Berry's phase~\cite{Berry} 
in a geometric engine as an extension of the Thouless pumping~\cite{Thouless,sinitsyn1,sinitsyn2}. 
We consider a small system sandwiched between two thermal reservoirs. 
If two parameters in the reservoirs and one parameter in the system Hamiltonian are controlled by an external agent, we can extract the work from the system. 
This is a natural application of the Thouless pumping~\cite{Thouless,sinitsyn1,sinitsyn2,ex-ch1,ex-ch2,ex-ch2.5,ex-ch3,ex-spin1,brouwer,ren,sagawa,yuge1,yuge2,watanabe,Hino-Hayakawa,Takahashi20JSP,FHHT} 
and geometric thermodynamics~\cite{Wang2021,Crooks,Bhandari20,Abiuso20,Brandner20,Hino2021,Hayakawa2021,Lu2022}.

\begin{figure}
\centering
\includegraphics[clip,width=6cm]{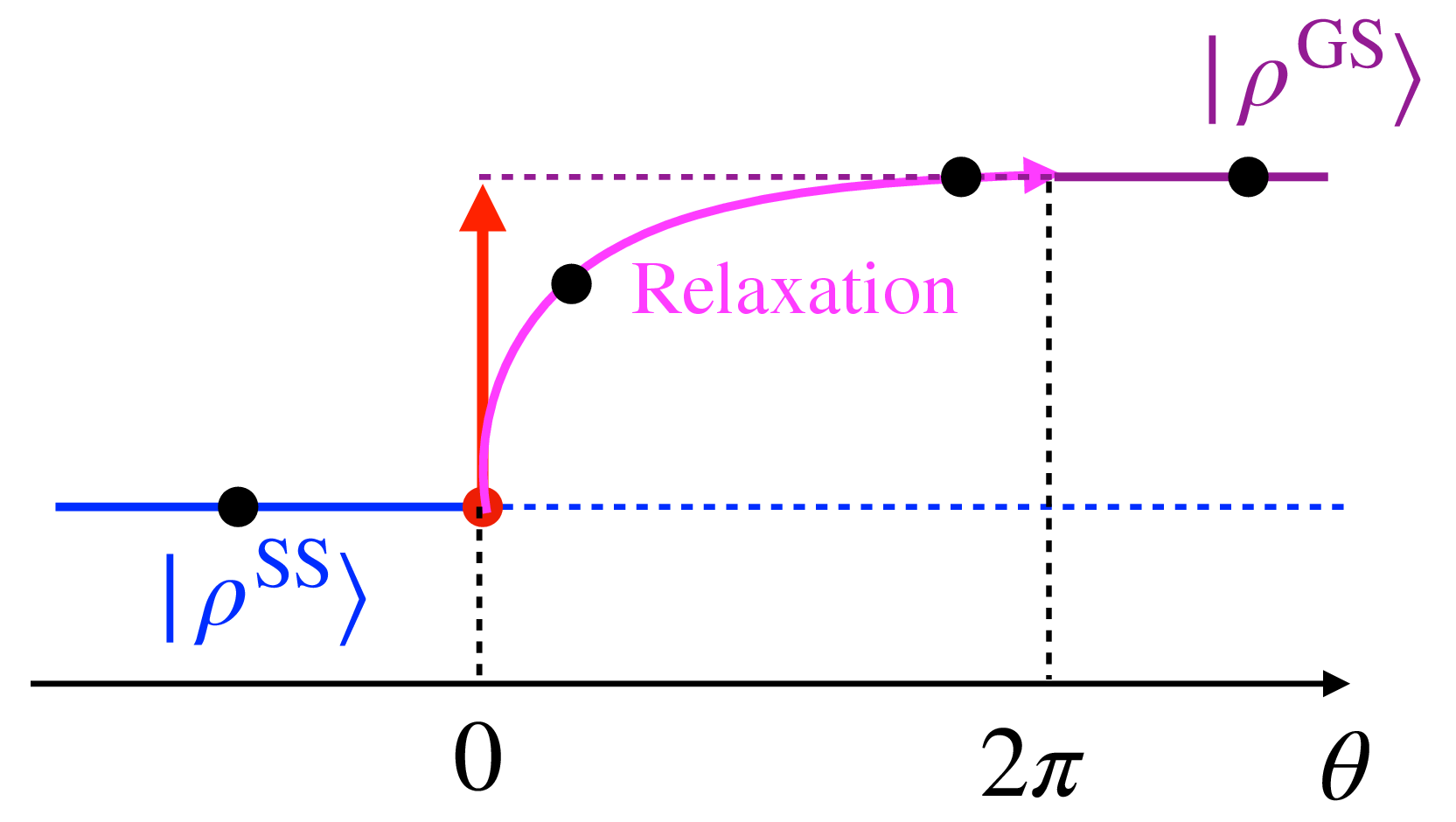}
\caption{A schematic of the process used for the implementation of the geometric demon. 
At $\theta=0$, we start modulating the parameters. 
Although the modulation is slow enough to prevent the system from being excited, 
the geometric term is suddenly added to the state vector. 
The state relaxes to the geometric state as time goes. 
The discrepancy between the steady state in the absence of the parameter modulation $|\hat\rho^{\rm SS}\rangle$ and the geometric state $|\hat\rho^{\rm GS}\rangle$ can be used to determine the work. 
}
\label{fig0}
\end{figure}

It is well known that the Kullback-Leibler (KL) divergence is non-negative, being zero only when the system is in a nonequilibrium steady state (NESS)~\cite{KL51,Sagawa20,Petz86,Petz03,Ruskai02,Hiai11,Hino2021,Hayakawa2021,Strasberg22}. 
It is also known that the KL divergence cannot increase for all completely positive and trace-preserving (CPTP) processes~\cite{KL51,Sagawa20,Petz86,Petz03,Ruskai02,Hiai11}. 
As a consequence, the KL divergence decreases monotonically towards zero in the relaxation process.
This monotonicity of the KL divergence corresponds to the second law of thermodynamics or the H-theorem.
Nevertheless, we must be careful in choosing the correct KL divergence in the presence of modulations. 
Indeed, we show that the correct choice of KL divergence under a periodic modulation is not the relative entropy between the time-evolving density matrix and that in a steady state but between the matrix and that in a geometric state because of the generation of a geometric term (see Fig.~\ref{fig0}).

We can therefore extract the work done by this geometric engine by means of an increase in the relative entropy corresponding to a negative entropy production. 
Let us call such an engine a geometric demon. 
The advantage of the geometric demon is that we do not need any costs for the feedback control,
 but we do need a non-adiabatic control of the system.

The organization of this paper is as follows. 
In Sec.\,\ref{sec:geometric_phase}, we explain the setup and the geometric state under the parameter modulation. 
We also present the corresponding KL divergences associated including the geometric contribution. 
In Sec.\,\ref{sec:WorkAndEfficiency}, we discuss the work, the heat, and the efficiency in the present setup. 
In Sec.\,\ref{sec:AndersonModel}, we apply our formulation to the Anderson model to demonstrate the protocol for realizing the geometric demon. 
In Sec.\,\ref{sec:ConcludingRemarks} we summarize our results and give their physical interpretations. 
We also give future perspectives. 
In Appendix \ref{appendix:initial}, we describe the derivation of the time evolution of the state including the geometric contribution. 
We also give a detailed expression for the BSN curvature. 
In Appendix \ref{Diff_App}, we explain the differences between the present study and the previous studies. 
In Appendix \ref{app:Anderson}, we present the detailed properties of the Anderson model. 
In Appendix \ref{App:Num_Th}, we present some detailed properties of the density matrix, the von Neumann entropy, and the heat flow from the comparison of the results of 
 the Born-like approximation used in this study with those obtained by the numerical solution.

\section{Geometric phase and entropy production}\label{sec:geometric_phase}

\begin{figure}
\centering
\includegraphics[clip,width=6cm]{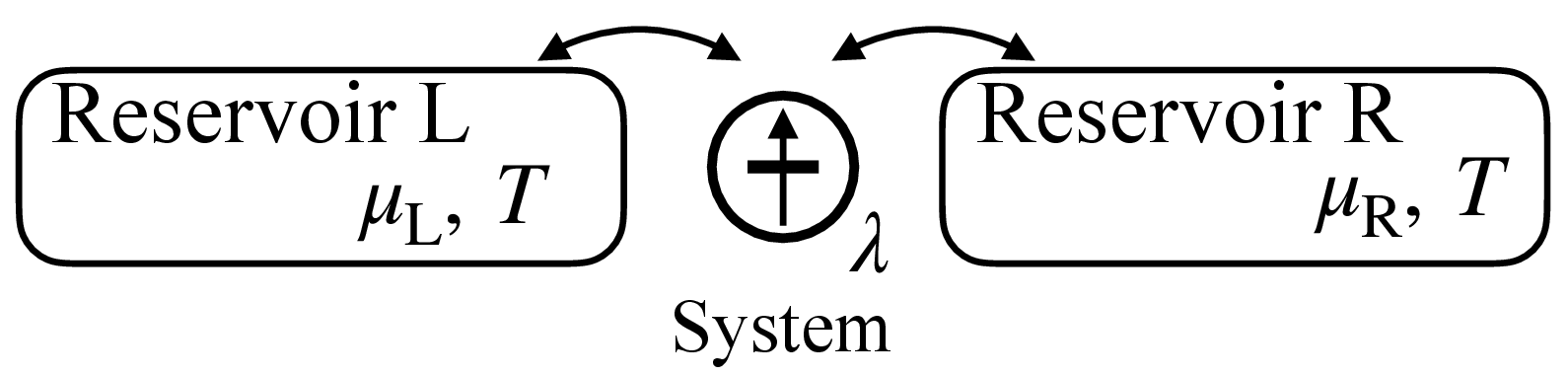}
\caption{A schematic of the model considered in the present paper. 
The system is connected to two reservoirs characterized by chemical potentials and temperatures. 
}
\label{model}
\end{figure}

In the present study, we focus on a system connected to two reservoirs depicted in Fig.\,\ref{model}.  
The left and right reservoirs are characterized by electrochemical potentials ($\mu_{\rm L}$ and $\mu_{\rm R}$) and temperature $T$, respectively.

\begin{figure}
\centering
\includegraphics[clip,width=6cm]{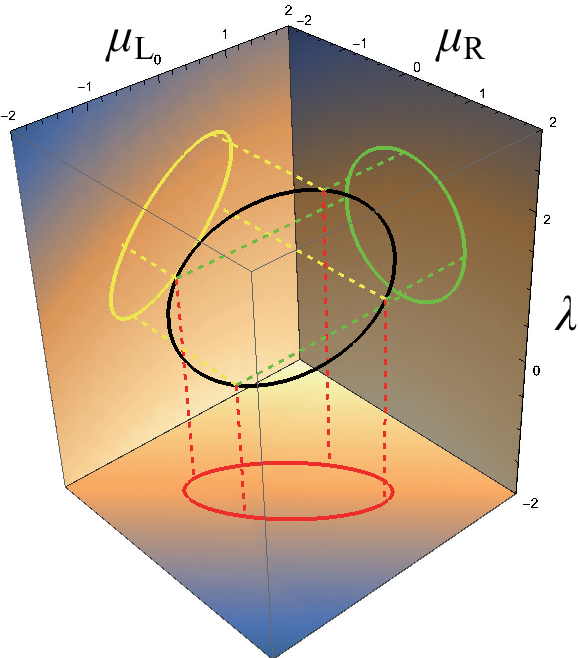}
\caption{A schematic of control parameters for $r=1$ and $\delta=\pi/4$. }
\label{fig_parameters}
\end{figure}

The electrochemical potentials are modulated as
\begin{equation}\label{chemi_pot}
\mu_{\mathrm{L}}=\overline{\mu}(1+r_{\mathrm{L}}\sin\theta),\quad
\mu_{\mathrm{R}}=\overline{\mu}[1+r_{\mathrm{R}}\sin(\theta+\delta)] ,
\end{equation}
where $\overline{\mu}:=\frac{1}{2\pi}\int_0^{2\pi}d\theta \mu_\alpha(\theta)$ is the one-cycle average of the electrochemical potential $\mu_\alpha$ in a reservoir $\alpha(=$${\mathrm{L}}$ or ${\mathrm{R}}$). 
We assume that $\mu_\alpha$ depends only on the modulation phase $\theta:=\omega t$, where $\omega$ and $t$ are the modulation angular frequency and time, respectively.
We also assume that the system Hamiltonian $\hat{H}(\lambda(\theta))$ is perfectly periodic in this paper, 
i.\,e., 
\begin{equation}
\hat{H}(\lambda(\theta))=\hat{H}(\lambda(\theta+2\pi))
\end{equation}
through a control parameter $\lambda(\theta)$, where $ \lambda(\theta):=1+r_H\cos\theta$.
In this paper, we consider only the case of $r:=r_\mathrm{L}=r_\mathrm{R}=r_H$ in order to reduce the number of parameters for simplicity.
To maintain the positivity of the parameters, we assume $|r|<1$. 
Thus, our system is characterized by a set of fixed parameters, such as $T$ and $\overline{\mu}$, and two control parameters, $r$ and $\delta$. 
To express a set of control parameters, 
we introduce 
\begin{equation}
\bm{\Lambda}(\theta,\delta):=\left(\lambda(\theta),\frac{\mu_{\rm L}(\theta)}{\overline{\mu}},\frac{\mu_{\rm R}(\theta,\delta)}{\overline{\mu}}\right)
\end{equation}
using $\Lambda_\mu$ as one of its components.
Figure~\ref{fig_parameters} shows a schematic of the control parameters for $r=1$ and $\delta=\pi/4$. 

We consider the master equation for the density matrix $\hat{\rho}(\theta,\delta)$:
\begin{align}\label{master}
\frac{d}{d\theta}|\hat\rho(\theta,\delta)\rangle=\epsilon^{-1} \hat K(\theta,\delta)|\hat\rho(\theta,\delta)\rangle, 
\end{align}
where $\hat{K}(\theta,\delta)$ is the evolution operator.
We use the vector notation $|\hat{\rho}(\theta,\delta) \rangle$ in Eq.\,\eqref{master}, where the components of the density matrix $\hat{\rho}(\theta,\delta)$ are aligned. 
We introduce the dimensionless parameter $\epsilon$ in Eq.\,\eqref{master} as
$\epsilon:=2\pi \omega/\Gamma$ where $\Gamma$ characterizes the hopping rate of the electrons from the reservoirs into the system.
Introducing the eigenvalue $\varepsilon_i$ with the subscript $i$ for $\hat{K}$, 
the corresponding left and right eigenstates $\langle \ell_i|$ and $|r_i\rangle$ satisfy the orthonormal relation $\langle \ell_i|r_j\rangle=\delta_{ij}$, 
if the eigenvalues are non-degenerate.
We assume that there exists a non-degenerate largest eigenvalue $\varepsilon_0:=0$ corresponding to a stationary state. 
Probability conservation leads to the left-zero eigenvector $\langle\ell_0|$, which is defined as $\langle\ell_0|\hat{K}=0$, whose diagonal components are $1$ in the matrix form and $0$ otherwise. 
The right zero eigenstate $|r_0\rangle$ satisfying $\hat{K}|r_0\rangle =0$ is also expressed as $|\hat{\rho}^{\rm SS}\rangle$ to specify the NESS.

In the absence of the modulations, the physical state is relaxed to $|\hat{\rho}^{\rm SS}\rangle$, since all eigenvalues except for $\varepsilon_0=0$ are negative. 
This suggests that the natural choice of the initial state would be $|\hat{\rho}^{\rm SS}\rangle$. 
Therefore, we analyze the case where the initial state is the steady state $|\hat{\rho}(0,\delta)\rangle=|r_0\rangle=|\hat{\rho}^{\mathrm{SS}}\rangle$. 

As soon as we start the modulation, the state relaxes to the other state, which we call the geometric state.  
In the presence of modulation, the state vector $|\hat{\rho}\rangle$ contains the contribution of a geometric phase.   
As shown in Appendix~\ref{general_initial}, the approximate expression for $|\hat{\rho}\rangle$ is given by 
\begin{align}
|\hat{\rho}(\theta,\delta)\rangle 
\simeq  |\hat{\rho}^{\rm SS}(\theta,\delta)\rangle+\sum_{i\neq0}^nC_i(\theta,\delta)|r_i (\theta,\delta)\rangle ,
\label{BSNConnection} 
\end{align}
where 
\begin{align}
C_i(\theta,\delta)&=- \int_0^{\theta}d\phi e^{\epsilon^{-1}\int^\theta_\phi dz \varepsilon_i(z,\delta)}\langle\ell_i(\phi,\delta) |\frac{d}{d\phi}|r_0(\phi,\delta)\rangle. 
\label{rhoCoeff}
\end{align}
The second term on the RHS of Eq.\,(\ref{BSNConnection}) is the Berry-Sinitsyn-Nemenman (BSN) connection~\cite{sinitsyn1,sinitsyn2}. 
A similar expression is derived in Ref.\,\cite{FHHT} 
( see Appendix \ref{Diff_App} for details of the relationship between our analysis and that in Ref.~\cite{FHHT}). 
Note that Eqs.~\eqref{BSNConnection} and \eqref{rhoCoeff} describe the evolution process from the initial steady state $|\hat{\rho}^\mathrm{SS}\rangle$, 
but the expression can be used if we are interested in the long time ($\theta\gg \epsilon$) behavior, where the choice of the initial condition is not important.
We also present the expression for $|\hat{\rho}(\theta,\delta)\rangle$ starting from a general initial condition in Appendix~\ref{general_initial} as well as the relaxation to Eqs.~\eqref{BSNConnection} and \eqref{rhoCoeff} for $\theta\gg \epsilon$.

The state described by Eqs.~\eqref{BSNConnection} and \eqref{rhoCoeff} relaxes to the geometric state:
\begin{align}
&|\hat{\rho}^{\rm{GS}}(\theta,\delta)\rangle:=\lim_{n\to\infty}\mathcal{T}\exp\left( \int^{\theta+2n\pi}_{0} d\phi \hat K(\phi,\delta)\right)|r_0(0,\delta)\rangle, 
\label{GeometricalState}
\end{align}
where we have introduced the time-ordering operator $\cal{T}$ defined as 
$ \mathcal{T}[\hat A(t_2)\hat B(t_1)\cdots \hat Z(t_n) ]:=\hat Z(t_n)\cdots \hat A(t_2)\hat B(t_1)$ with $t_1<t_2<\cdots <t_n$.
Since this relaxation process is fast, 
we can use the approximate expression for the geometric state as
\begin{equation}\label{GS}
|\hat{\rho}^{\rm{GS}}(\theta,\delta)\rangle\simeq \mathcal{T}\exp\left( \int^{\theta+2\pi}_{0} d\phi \hat K(\phi,\delta)\right)|r_0(0,\delta)\rangle, 
\end{equation}
Note that the trace preservation is always satisfied for an arbitrary $\theta$ from $\langle\ell_0|r_i\rangle=\delta_{0i}$ and $\langle\ell_0|\hat K=0$ in the dynamics described by Eq.~\eqref{master}.
In this paper, we restrict our interest to quasi-adiabatic processes. 
There is an analogy between Eq.\,\eqref{BSNConnection} and the path-integral with the Born approximation in the quantum theory. 
The phase factor $e^{\epsilon^{-1}\int^\theta_\phi dz \varepsilon_i(z,\delta)}$ in Eq.\,\eqref{rhoCoeff} can be interpreted as free relaxation instead of free propagation. 
If the modulation is not slow, i.e.\, $\epsilon$ is larger than $|\varepsilon_{i\neq 0}|$, it would not be adequate to ignore the higher order processes.
The expressions in Eqs.\,\eqref{BSNConnection} and \eqref{rhoCoeff} are compatible with the slow-modulation approximation employed in Ref.\,\cite{Hayakawa2021}. 
The leading contribution of the modulation to the entropy production is indicated by the second term on the right-hand side (RHS) of Eq.\,(\ref{BSNConnection}).

As shown in Appendix~\ref{BSN_Suppl}, 
the deviation from the initial state after a modulation cycle satisfying $|r_i(2\pi,\delta)\rangle=|r_i(0,\delta)\rangle$ is written as
\begin{align}
&\Delta |\hat{\rho}\rangle:=
|\hat{\rho}(2\pi,\delta)\rangle-|\hat{\rho}(0,\delta)\rangle=\sum_{i\neq0}\mathcal{C}_i(\delta)|r_i (0,\delta)\rangle  \label{BSNConnection3} 
\end{align}
with
\begin{align}
&\mathcal{C}_i(\delta):=\int_0^{2\pi} d\phi e^{\epsilon^{-1}\int^{2\pi}_\phi dz \varepsilon_i(z,\delta)}\mathcal{A}^\mu_i \frac{\partial \Lambda_\mu}{\partial \phi} ,
\label{BSNCoeffOC}
\end{align} 
where we have introduced the BSN connection $\mathcal{A}^\mu_i$ as
\begin{align}
\mathcal{A}^\mu_i(\phi,\delta):=-\langle\ell_i(\Lambda(\phi,\delta)) |\frac{\partial}{\partial \Lambda_\mu}|r_0(\Lambda(\phi,\delta))\rangle .
\label{DefBSNConnection}
\end{align}
According to $\mathcal{A}^\mu_i$, we define the BSN curvature as
\begin{align}
F^{\mu\nu}_i(\theta,\delta):
=\left(\frac{\partial\mathcal{A}^\nu_i}{\partial \Lambda_{\mu}}\right)_\theta-\left(\frac{\partial\mathcal{A}^\mu_i}{\partial \Lambda_{\nu}}\right)_\theta. 
\label{DefBSNCurvaure}
\end{align}
Due to the damping factor in Eq.\,(\ref{BSNCoeffOC}), 
the contribution of the BSN curvature is localized in $\theta$.

Now, let us consider the KL divergence. 
The proper KL divergence is not $D^\mathrm{KL}(\hat{\rho}(\theta,\delta)||\hat{\rho}^\mathrm{SS}(\theta,\delta))$ but $D^\mathrm{KL}(\hat{\rho}(\theta,\delta)||\hat{\rho}^\mathrm{GS}(\theta,\delta))$, 
where we have defined the KL divergence~\cite{KL51,Sagawa20,Petz86,Petz03,Ruskai02,Hiai11,Hino2021,Hayakawa2021} as
\begin{align}\label{KLDef}
D^{\mathrm{KL}}(\hat\rho||\hat\sigma):
=&\mathrm{Tr}\left[\hat\rho (\ln \hat\rho -\ln \hat\sigma)\right] .
\end{align}  
Since we expect $|\hat \rho(\theta,\delta)\rangle$ to quickly relax to $|\hat \rho^{\rm GS}(\theta,\delta)\rangle$, the KL divergence approximately satisfies 
\begin{equation}
D^\mathrm{KL}(\hat{\rho}(\theta,\delta)||\hat{\rho}^\mathrm{GS}(\theta,\delta))\simeq 0
\end{equation}
for $\theta\ge 2\pi$.
The initially positive $D^\mathrm{KL}(\hat{\rho}(\theta,\delta)||\hat{\rho}^\mathrm{GS}(\theta,\delta))$ decreases monotonically as $\theta$ approaches zero.
This is the H-theorem.

The production of the total entropy associated with the KL divergence should be  
\begin{align}\label{KL-heat} 
& \Delta \Sigma(\theta,\delta) \notag\\
& \quad :=D^\mathrm{KL}(\hat{\rho}(0,\delta)||\hat{\rho}^\mathrm{GS}(0,\delta))- D^\mathrm{KL}(\hat{\rho}(\theta,\delta)||\hat{\rho}^\mathrm{GS}(\theta,\delta)).
\end{align}
This $\Delta \Sigma(\theta,\delta)$ which is always non-negative could play an important role.

On the other hand, we can consider the entropy production during the one-cycle modulation starting from the NESS at $\theta=0$ as
\begin{align}\label{DeltaSDef}
\Delta S(\delta):
&= D^{\mathrm{KL}}(\hat\rho(0,\delta)||\hat\rho^{\mathrm{SS}}(0,\delta))-D^{\mathrm{KL}}(\hat\rho(2\pi,\delta)||\hat\rho^{\mathrm{SS}}(0,\delta)) 
\notag\\
&=-D^{\mathrm{KL}}(\hat\rho(2\pi,\delta)||\hat\rho^{\mathrm{SS}}(0,\delta))
.
\end{align}
This $\Delta S(\delta)$ is always non-positive due to the non-negativity of the KL divergence. 
We could expect $\Delta S(\delta)\ge 0$,  
since the KL divergence is expected to decrease with $\theta$~\cite{KL51,Sagawa20,Petz86,Petz03,Ruskai02,Hiai11} as 
\begin{equation}\label{DS>=0}
\Delta S(\delta)=-\int_0^{2\pi}d\theta \frac{\partial}{\partial \theta}D^{\rm KL}(\hat{\rho}(\theta,\delta)||\hat{\rho}^\mathrm{SS}(\theta,\delta))
\ge 0.
\end{equation}
However, the compatible value of $\Delta S(\delta)$ in Eq.~\eqref{DS>=0} with the non-negative KL divergence is zero under the periodic modulation.
Nevertheless, $\Delta S(\delta)$ can be negative for most $\delta$ as will be shown, because $|\hat{\rho}(\theta,\delta)\rangle$ for $\theta\ne 0$ is not described by $|\hat{\rho}^\mathrm{SS}(0,\delta)\rangle$ but by Eqs.~\eqref{BSNConnection} and \eqref{rhoCoeff}. 
This implies that $D^{\rm KL}(\hat{\rho}(\theta,\delta)||\hat{\rho}^\mathrm{SS}(\theta,\delta))$ is not proper KL divergence.
In fact, as will be shown, $D^{\rm KL}(\hat{\rho}(\theta,\delta)||\hat{\rho}^\mathrm{SS}(\theta,\delta))$ is not monotonic with $\theta$ but oscillates with $\theta$.
Thus, we can extract the work from the {\it negative entropy production} $\Delta S(\delta)<0$.

\section{Work and efficiency}\label{sec:WorkAndEfficiency}

\subsection{First law of thermocynamics}

Now, we discuss the thermodynamic relations to construct the geometric demon. 
First, we consider the first law of thermodynamics~\cite{Strasberg22,Benenti,Kurizki}. 
The internal energy is given by 
\begin{equation}
E(\theta, \delta)=\mathrm{Tr}[\hat H(\lambda(\theta))\hat\rho(\theta,\delta)]. 
\end{equation}
The rate of change of internal energy is given by 
\begin{align}
\frac{\partial E}{\partial\theta}(\theta, \delta)
=&\mathrm{Tr}\left[\frac{\partial\hat H(\lambda(\theta))}{\partial\theta}\hat \rho(\theta,\delta)\right] 
+\mathrm{Tr}\left[\hat H(\lambda(\theta))\frac{\partial\hat \rho(\theta,\delta)}{\partial\theta}\right]. 
\label{dE}
\end{align}
Here, the first term on the RHS of Eq.\,\eqref{dE} corresponds to the power, which can be rewritten as
\begin{align}
\label{Work_rate}
&\mathscr{P}(\theta,\delta):=
 {\rm Tr} 
 \left[
  \hat{\rho}(\theta,\delta)\frac{ \partial \hat{H}(\lambda(\theta))}{\partial \lambda(\theta)}  \right] 
  \dot{\lambda}(\theta) ,
\end{align}
where $\dot{\lambda}(\theta):=\frac{d}{d\theta}\lambda(\theta)$.
Thus the work generated from the initial state to $\theta$ becomes 
\begin{align}
\label{Work_rate}
&\mathcal{W}(\theta,\delta):=
\int_0^{\theta} d\phi 
 \mathscr{P}(\phi,\delta).
\end{align}
By definition, the relation $ \mathcal{\dot W}(\theta,\delta)=\mathscr{P}(\theta,\delta)$ holds. 
The work $W(\delta)$ during the one cycle is defined as 
\begin{align}\label{W}
&W(\delta) :=  
\mathcal{W}(2\pi,\delta). 
\end{align}
The work $W(\delta)$ and power $\mathscr{P}(\theta,\delta)$ can be positive or negative depending on the situation. 
A positive $\mathscr{P}(\theta,\delta)$ is interpreted as a power input from the external agent, 
while a negative $\mathscr{P}(\theta,\delta)$ can be interpreted as a power loss. 
The second term on the RHS of Eq.\,\eqref{dE} corresponds to the heat flow 
\begin{align}
\label{HeatFlow}
&\mathcal{\dot Q}(\theta,\delta):=
 {\rm Tr} 
 \left[
  \frac{\partial \hat{\rho}(\theta,\delta)}{\partial \theta} \hat{H}(\lambda(\theta)) \right].
\end{align}
Thus the heat generated from the initial state to $\theta$ becomes 
\begin{align}
&\mathcal{Q}(\theta,\delta) :=  
\int^{\theta}_{0}d\phi  \mathcal{\dot Q}(\phi,\delta).  
\end{align}
The heat generated during one cycle is defined as 
\begin{align}\label{Q}
&Q(\delta) :=  
\mathcal{Q}(2\pi,\delta).  
\end{align}

Thus, we can rewrite Eq.\,\eqref{dE} in the well-known form of the first law of thermodynamics as 
\begin{align}
\frac{\partial}{\partial \theta}E(\theta,\delta)=\frac{\partial}{\partial \theta}\mathcal{W}(\theta,\delta)+\frac{\partial}{\partial \theta}\mathcal{Q}(\theta,\delta).
\end{align} 
The power $\mathscr{P}(\theta,\delta)$, the work $W(\delta)$, the heat flow $\mathcal{Q}(\theta,\delta)$, and the heat $Q(\delta)$ can be calculated once we know the density matrix $\hat{\rho}(\theta,\delta)$ as in Eqs.\,\eqref{BSNConnection} and \eqref{rhoCoeff}.

\subsection{Second law of thermocynamics}

Because the KL divergence satisfies the second law of thermodynamics or the H theorem expressed as $D^\mathrm{KL}(\hat{\rho}(\theta,\delta)||\hat{\rho}^\mathrm{GS}(\theta,\delta))\ge 0$ and its monotonicity, 
we can construct the thermodynamics for the driven system by considering the system entropy and the heat between the system and the reservoirs.
As will be shown, the system entropy differs from the von Neumann entropy due to the contribution of the driven terms.

Let us consider $\Delta\Sigma(\theta,\delta)$ defined as
\begin{equation}
\Delta\Sigma(\theta,\delta):=D^\mathrm{KL}(\hat{\rho}(0,\delta)||\hat{\rho}^\mathrm{GS}(0,\delta))-D^\mathrm{KL}(\hat{\rho}(\theta,\delta)||\hat{\rho}^\mathrm{GS}(\theta,\delta))
\end{equation}
which can be rewritten as $\Delta\Sigma(\theta,\delta)=\Delta S_{\rm sys}+\Delta S_{\rm r}$ where $\Delta S_{\rm sys}$ and $\Delta S_{\rm r}$ express the changes of the system entropy and the reservoir, respectively. 
Because $\Delta S_{\rm r}$ is expressed as $\Delta S_{\rm r}=-\beta \mathcal{Q}$, we obtain~\cite{Strasberg22} 
\begin{align}
\label{1st_law_th}
\Delta\Sigma(\theta,\delta)
=\Delta S_{\rm sys}(\theta,\delta)-\beta \mathcal{Q}(\theta,\delta), 
\end{align}
where $\beta:=1/T$, $\Delta S_{\rm sys}(\theta,\delta):=S_{\rm sys}(\theta,\delta)+\mathrm{Tr} [\hat{\rho}(0,\delta) \ln \hat{\rho}(0,\delta)]$ with
\begin{align}\label{von-Neumann}
S_{\rm sys}(\theta,\delta):=&
S^\mathrm{vN}(\theta,\delta)
+\beta \mathcal{Q}(\theta,\delta)
+\beta \mathcal{\tilde Q}(\theta,\delta),  
\end{align}
where we have introduced von Neumann entropy
\begin{align}\label{von-Neumann}
S^\mathrm{vN}(\theta,\delta):&=-\mathrm{Tr} [\hat{\rho}(\theta,\delta) \ln \hat{\rho}(\theta,\delta)] ,
\end{align}
and
\begin{align}
\beta \mathcal{\tilde Q}(\theta,\delta):&=\mathrm{Tr} [\hat{\rho}(\theta,\delta) \ln \hat{\rho}^{\rm GS}(\theta,\delta)]
-\mathrm{Tr} [\hat{\rho}(0,\delta) \ln \hat{\rho}^{\rm GS}(0,\delta)] .
\label{tildeQ}
\end{align}
This $S_{\rm sys}(\theta,\delta)$ is the system entropy which is deviated from von Neumann entropy $S^\mathrm{vN}(\theta,\delta)$.
Note that the $S_{\rm sys}(\theta,\delta)$ is reduced to $S^\mathrm{vN}(\theta,\delta)$ when the parameter in the Hamiltonian is not modulated and the system is relaxed to an equilibrium state.
Since $\hat \rho^{\rm GS}(\theta,\delta)$ is reduced to be $\hat \rho^{\rm eq}(\theta,\delta)\propto e^{-\beta \hat H}$ in such a situation, 
the RHS of Eq.\,\eqref{tildeQ} becomes $\int_0^\theta d\phi \frac{d}{d\phi}[-\beta \hat H \rho]= -\beta \int_0^\theta d\phi [\hat H \dot \rho]=-\beta \mathcal{Q}(\theta,\delta)$. 
Since we have already obtained the expression of $\hat{\rho}(\theta,\delta)$ as in Eqs.~\eqref{BSNConnection} and \eqref{rhoCoeff}, one can obtain the heat $\mathcal{Q}(\theta,\delta)$. 

\subsection{Absorbing heat, work, and efficiency}

Because $\dot{\mathcal{Q}}(\theta,\delta)$ can be negative, the heat $\mathcal{Q}(\theta,\delta)$ can be negative.
We can decompose $\mathcal{Q}(\theta,\delta)$ into the absorbing heat and the releasing heat as  
\begin{align}\label{defW_g}
Q_{A}(\delta):&=\int_0^{2\pi}d\theta \dot{\mathcal{Q}}(\theta,\delta)\Theta(\dot{\mathcal{Q}}(\theta,\delta)),
\\
Q_{R}(\delta):&=Q(\delta)-Q_{A}(\delta),
\end{align}
where $\Theta(x)$ is the Heaviside step function that satisfies $\Theta(x)=1$ if $x\ge 0$ and $\Theta(x)=0$ otherwise. 
If we consider a slow modulation process, we expect the deviation of the state from the steady state $\hat \rho^{\rm SS}$ to be small 
and the time evolution will be nearly periodic. 
Thus, $\Delta E(\delta):=E(2\pi,\delta)-E(0, \delta)$ would be much smaller than the other variables such as $W(\delta)$ and $Q(\delta)$. 
(Namely, we expect $|W(\delta)|, |Q(\delta)|\gg |\Delta E(\delta)|$). 
This implies that $Q_A(\delta) \ge |Q(\delta)|\simeq  |W(\delta)| \ge W(\delta)$  holds due to the first law of thermodynamics. 
We can prove this relation by considering a slow modulation process $\epsilon\ll 1$ under the rotational wave approximation~\cite{Breuer}.
\footnote{
We estimate $\Delta E(\delta)$ as follows. 
$\Delta E(\delta)={\rm Tr}[\hat H \hat\rho(2\pi,\delta)]-{\rm Tr}[\hat H \hat\rho(0,\delta)]=\sum_{i,j}[\hat H_{ij} \hat\rho^{\rm GS}_{ji}-\hat H_{ij} \hat\rho^{\rm SS}_{ji}]$. Since both $\hat\rho^{\rm GS}_{ij}$ and $\hat\rho^{\rm SS}_{ij}$ are diagonal under the rotational wave approximation \cite{Breuer}, $\Delta E(\delta)$ is estimated as
$\Delta E(\delta)=\sum_{i}[\hat H_{ii} \hat\rho^{\rm GS}_{ii}-\hat H_{ii} \hat\rho^{\rm SS}_{ii}]=\sum_{i}[\hat H_{ii} (\hat\rho^{\rm SS}_{ii}+C_1r_{1,i}+\cdots)-\hat H_{ij} \hat\rho^{\rm SS}_{ii}]\simeq C_1\sum_{i}[\hat H_{ii} r_{1,i}]$,
 where $|r_1\rangle$ is the right eigenvector of $\hat{K}$ associated with the slowest eigenmode.
$\dot{\mathcal{Q}}={\rm Tr}[\hat H \dot{\hat\rho}]={\rm Tr}[\hat H \epsilon^{-1}\mathcal{K} \hat\rho]\simeq C_1{\rm Tr}[\hat H \epsilon^{-1}\varepsilon_1 \hat{r}_1]=C_1\epsilon^{-1}\varepsilon_1\sum_i[\hat H_{ii}r_{1,i}]$, and
$\mathscr{P}=\int d\theta {\rm Tr}[\hat H \dot{\hat\rho}]=\epsilon^{-1}\int d\theta C_1\varepsilon_1\sum_i[\hat H_{ii}r_{1,i}]$, where $\hat{r}_1$ is the matrix expression of $|r_1\rangle$. 
Thus, $\dot{\mathcal{Q}}$ and $\mathscr{P}$ are much larger than $\Delta E$ because of the factor $\epsilon^{-1}$ in the quasi-adiabatic situation.
 }
If the work $W(\delta)$ is negative, the system can be regarded as an engine in which the work done by the system is larger than the work done by the reservoirs.

Thus, to construct a geometric demon, we require that $W(\delta)<0$ to use the negative entropy production $\Delta S(\delta)<0$. 
To the best of our knowledge, we cannot determine the sign of $W(\delta)$ in general.
Therefore, we use the Anderson model to show that $W(\delta)$ can be negative. 

Using the absorbed energy, the work, and the efficiency can be introduced as  
\begin{equation}
\label{efficiency}
\eta(\delta):=-\frac{W(\delta)}{Q_{\mathrm{A}}(\delta)}.    
\end{equation}
When the work becomes positive, the efficiency becomes ill-defined.

\section{Application to the Anderson model}\label{sec:AndersonModel}

Let us apply the general framework to the Anderson model for a quantum dot (QD), 
where a single dot is coupled to two electron reservoirs. 
\footnote{Although higher energy levels exist in realistic quantum dot systems, for simplicity here we consider only the single energy level for simplicity.}
The total Hamiltonian $\hat{H}^{\rm tot}$ can be written as 
\begin{align}\label{H_tot}
\hat{H}^{\rm tot}:&=\hat{H}+\hat{H}^{\rm r}+\hat{H}^{\rm int} ,
\end{align}
where $\hat{H}$, the reservoir Hamiltonian $\hat{H}^{\rm r}$ and the interaction Hamiltonian $\hat{H}^{\rm int}$ are, respectively, given by
\begin{align}\label{H_s}
\hat{H}&=\sum_\sigma \epsilon_0 \hat{d}_\sigma^\dagger \hat{d}_\sigma+U(\theta) \hat{n}_\uparrow \hat{n}_\downarrow ,
\\
\label{H_bath}
\hat{H}^{\rm r}&=\sum_{\alpha,k,\sigma}\epsilon_k \hat{a}_{\alpha,k,\sigma}^\dagger \hat{a}_{\alpha,k,\sigma} ,
\\
\hat{H}^{\rm int}&=\sum_{\alpha,k,\sigma}V_\alpha \hat{d}_\sigma^\dagger \hat{a}_{\alpha,k,\sigma}+{\rm h.c.},
\end{align}
where $\hat{a}_{\alpha, k,\sigma}^\dagger$ and $\hat{a}_{\alpha, k,\sigma}$ are 
the creation and annihilation operators, respectively, for the electrons in the reservoirs $\alpha(=$L or R) with the wavenumber $k$, 
energy $\epsilon_k$, and spin $\sigma(=\uparrow$ or $\downarrow$).
Furthermore, $\hat{d}^\dagger_\sigma$ and $\hat{d}_\sigma$ are those in the QD, and $\hat{n}_{\sigma}=\hat{d}^\dagger_{\sigma}\hat{d}_{\sigma}$. 
$U(\theta):=U_0\lambda(\theta)$ and $V_\alpha$ are, respectively, 
the time-dependent electron-electron interaction in the QD and the transfer energy between the QD and the reservoir $\alpha$. 
We adopt a model in the wide-band limit for the reservoirs.
We denote the line width in this paper as $\Gamma=\pi \varrho (V_\mathrm{L}^2+V_\mathrm{R}^2)$, 
where $\varrho$ is the density of states in the reservoirs. 
For simplicity, we set $V_L=V_R$.

The Anderson model for the QD has four states: doubly occupied, singly occupied with an up-spin, singly occupied with a down-spin, and empty.
Therefore, the density matrix is expressed as a $4\times 4$ matrix.
However, as shown in Ref.\,\cite{Hayakawa2021}, $\hat{\rho}(\theta,\delta)$ of the Anderson model under the wide-band approximation is reduced to a diagonal matrix, 
where the diagonal elements correspond to the probability of finding the states in the empty state $\rho_e$, 
the down-spin state $\rho_{\downarrow}$, the up-spin state $\rho_\uparrow$, and the doubly occupied state $\rho_d$, respectively. 
This implies that the model is quasi-classical.

The trace-preserving condition $\mathrm{Tr}\hat\rho=\rho_e+\rho_\uparrow+\rho_\downarrow+\rho_d=1$ reduces to the probability preserving condition. 
The explicit forms of the evolution matrix $\hat{K}$ and the corresponding eigenstates $\langle \ell_i|$ and $|r_i\rangle$ are summarized in Ref.~\cite{Hayakawa2021}. 
The explicit forms of the BSN connection in Eq.~\eqref{BSNCoeffOC} are presented in Appendix~\ref{app:Anderson}.

Figure~\ref{fig:density_matrix} shows the time evolution of the elements $\rho_{e}$, $\rho_{\downarrow}$, $\rho_{\uparrow}$ and $\rho_{d}$ of $\hat{\rho}(\theta)$
(see also Appendix~\ref{app:Anderson}).
This figure clearly supports the positivity of all elements, therefore the evolution dynamics satisfies CPTP properties.
As shown in Appendix~\ref{App:Num_Th} the approximate expression of the density matrix in Eqs.~\eqref{BSNConnection} and \eqref{rhoCoeff} is very precise, where the deviation between the numerical solution of Eq.~\eqref{master} and the approximate expression in Eqs.~\eqref{BSNConnection} and \eqref{rhoCoeff} is almost invisible,  


\begin{figure}
\centering
\includegraphics[clip,width=6cm]{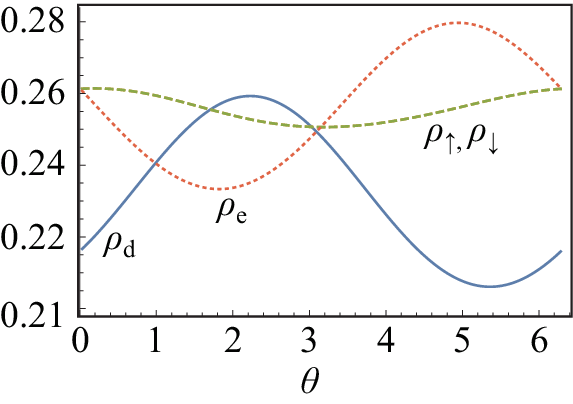}
\caption{Scaled time ($\theta$) evolutions of the elements of the density matrix. 
The parameters are set to be $\beta \bar\mu=0.1$, $\beta U_0=0.1$, and $\beta \epsilon_0=0.1$. 
}
\label{fig:density_matrix}
\end{figure}

\begin{figure} 
\centering
\includegraphics[clip,width=8cm]{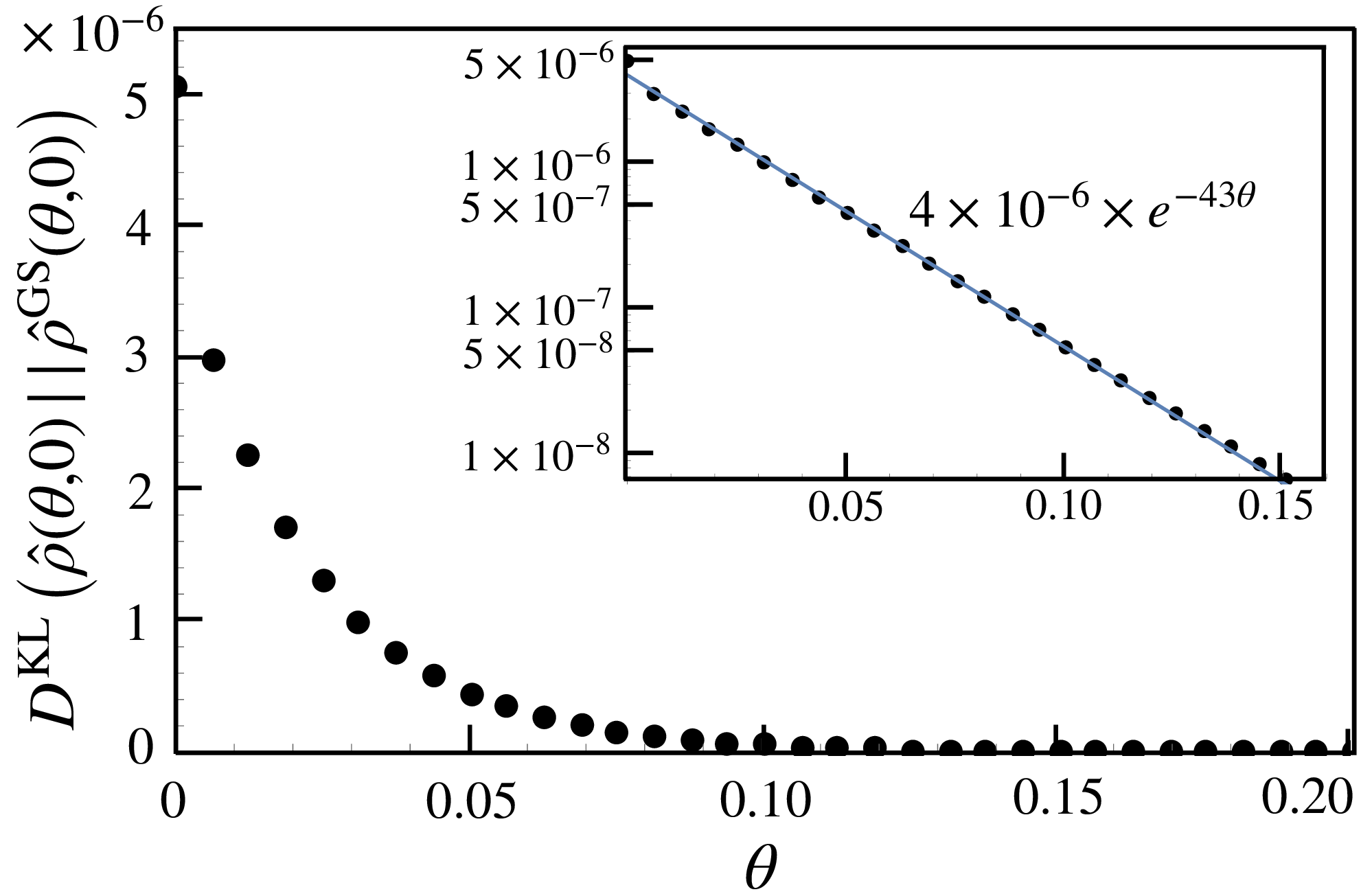}
\caption{Plot of $D^\mathrm{KL}(\hat{\rho}(\theta,0)||\hat{\rho}^\mathrm{GS}(\theta,0))$ against $\theta$ for $\beta U_0=0.1$ and $r=0.9$. 
The inset is the semi-log plot of $D^\mathrm{KL}(\hat{\rho}(\theta,\delta)||\hat{\rho}^\mathrm{GS}(\theta,\delta))$ against $\theta$.
}
\label{figS1}
\end{figure}

In the following, we set $\epsilon=0.1$ to get some explicit results for the Anderson model.
Converting $|\hat \rho(\theta)\rangle$ to the matrix form $\hat\rho(\theta,\delta)$ and inserting it into Eq.\,(\ref{DeltaSDef}), 
we get $D^\mathrm{KL}(\hat{\rho}(\theta,\delta)||\hat{\rho}^\mathrm{GS}(\theta,\delta))$. 
The time evolution of $D^\mathrm{KL}(\hat{\rho}(\theta,0)||\hat{\rho}^\mathrm{GS}(\theta,0))$ is shown in Fig.\,\ref{figS1} for $r=0.9$, and $\beta U_0=0.1$. 
This figure clearly supports the H-theorem, where the KL divergence decreases monotonically as an exponential function of $\theta$ to approach zero. 
Thus, we can evaluate $\mathcal{Q}_\mathrm{KL}(\delta)$ introduced in Eq.~\eqref{KL-heat}. 
We have also confirmed $\hat{\rho}(\theta=2\pi,\delta)=\hat{\rho}(4\pi,\delta)$ within the numerical accuracy.
This justifies the practical treatment of Eq.~\eqref{GS}.
As a result, the trivial relation $D^\mathrm{KL}(\hat{\rho}(\theta,\delta)||\hat{\rho}^\mathrm{GS}(\theta,\delta))= 0$ holds for $\theta\ge 2\pi$.

\begin{figure}
\centering
\includegraphics[clip,width=7cm]{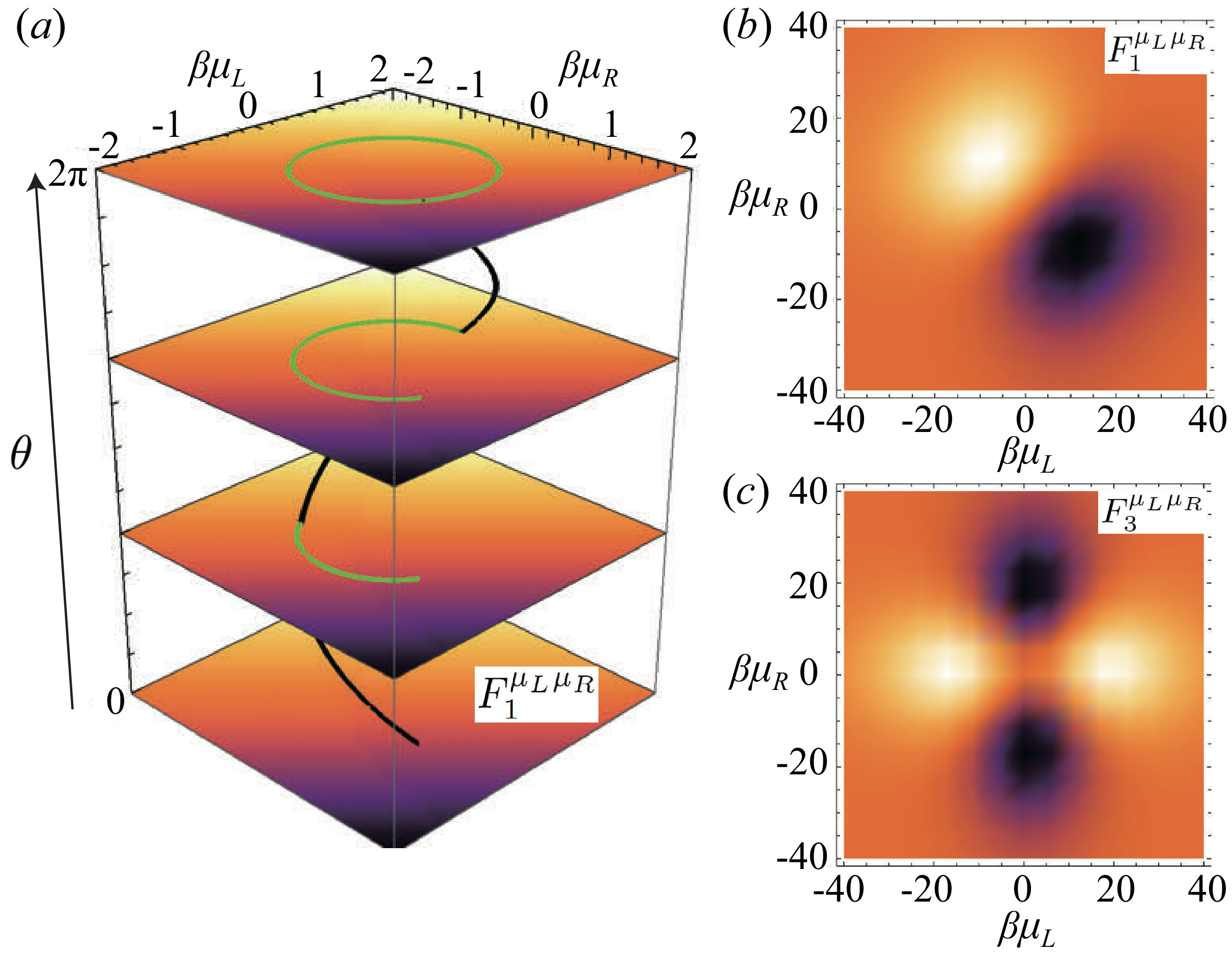}
\caption{(a)\,Schematics of a contour of the integral of $\mathcal{C}_1$, where the black solid line is the trajectory of the parameters. 
The color scale at a value of $\theta$ expresses $F^{{\mu_\mathrm{L}}{\mu_\mathrm{R}}}_1$. 
(b) and (c) The BSN curvatures $F^{{\mu_\mathrm{L}}{\mu_\mathrm{R}}}_1$ and $F^{{\mu_\mathrm{L}}{\mu_\mathrm{R}}}_3$ at $\theta= 2\pi$ are plotted. 
The parameters are set to be $\beta\bar\mu=0.1$, $\beta U_0=0.1$, and $\beta \epsilon_0=0.1$ for all figures. }
\label{fig3}
\end{figure}

Figure \ref{fig3}(a) shows the contour of the integral given in Eq.\,(\ref{BSNCoeffOC}) in the parameter space $(\beta\mu_{\rm L}(\theta),\beta\mu_{\rm R}(\theta,\delta),\theta)$. 
For the explicit calculation, we use Eqs.\,\eqref{BSNConnection} and \eqref{rhoCoeff} with $\theta=2\pi$ for Figs.~\ref{fig3} (b) and (c). 
As can be seen, the BSN curvature is always present for $\theta>0$, although its magnitude decreases with $\theta$. 
The BSN curvatures at specific $\theta$s are plotted in Figs.\,\ref{fig3}(b) and (c), 
where the peak and the dip of $F^{\mu_\mathrm{L}\mu_\mathrm{R}}_1$ are approximately located at $\beta \mu_\mathrm{R}=-\beta\mu_\mathrm{L} \approx 10$, $\beta \mu_\mathrm{R}=-\beta\mu_\mathrm{L} \approx -10$, respectively,  
while the peak and the dip of $F^{\mu_\mathrm{L}\mu_\mathrm{R}}_3$ are approximately located at $\beta \mu_\mathrm{L} \approx \pm 10, \beta\mu_\mathrm{R}=0$ and $\beta \mu_\mathrm{R} \approx \pm 10, \beta\mu_\mathrm{L}=0$, respectively.
Integrating by parts, we can verify that $C_2(\theta,\delta)=0$ and $F^{\mu_\mathrm{L}\mu_\mathrm{R}}_2=0$, 
because $\langle \ell_2|$ is independent of $\theta$.

\begin{figure}
\centering
\includegraphics[clip,width=6cm]{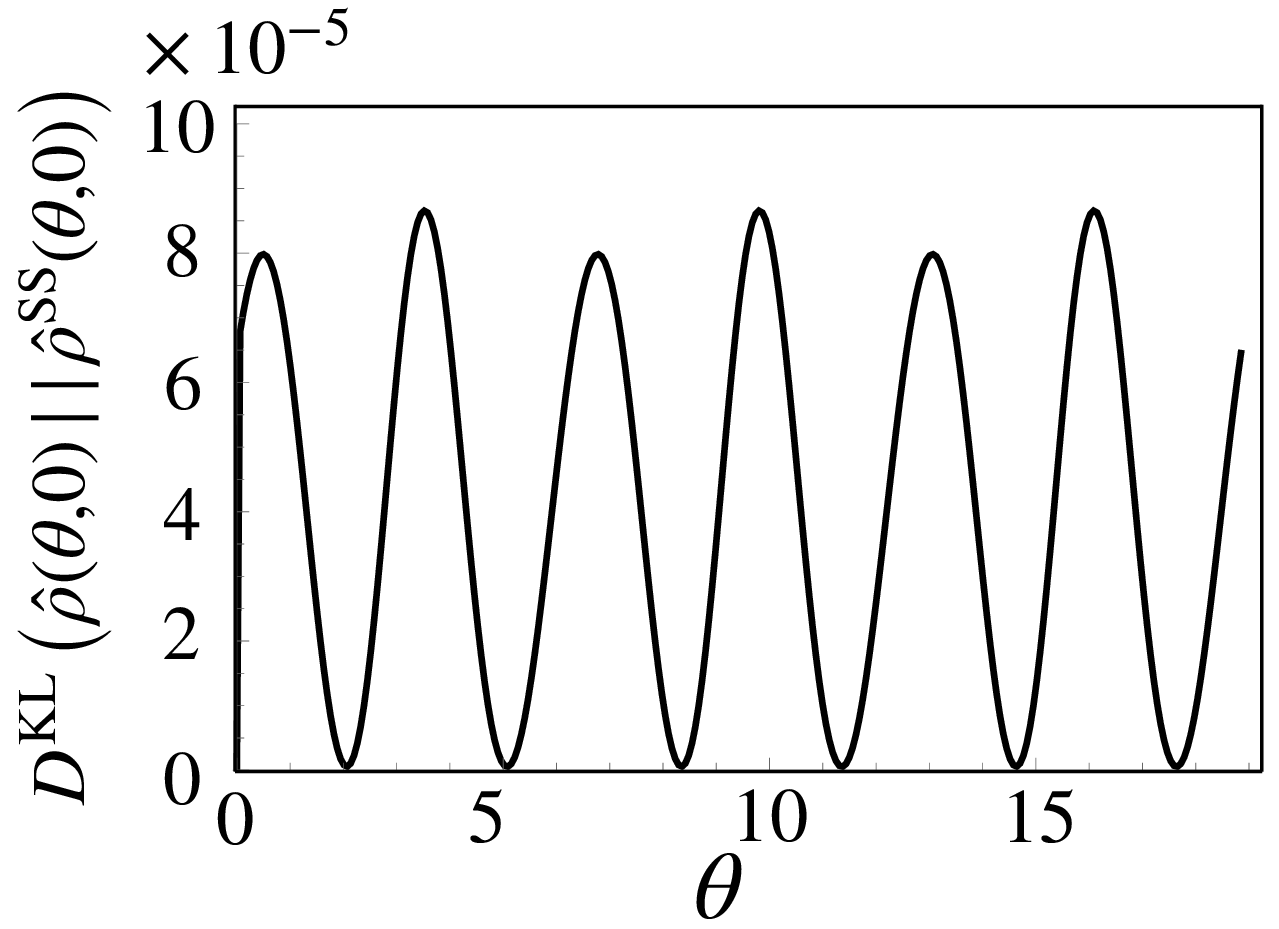}
\caption{Plots of $D^\mathrm{KL}(\rho(\theta,0)|\hat{\rho}^{\rm SS}(\hat{\rho}(\theta,0))$ versus $\theta$ for $r=0.9$ and $\beta U_0=0.3$.}
\label{DKL}
\end{figure}

Figure \ref{DKL} plots $D^\mathrm{KL}(\hat{\rho}(\theta,0)||\hat{\rho}
^{\rm SS}(\rho(\theta,0))$ versus $\theta$.  
This illustrates the rapid relaxation near $\theta=0$ and the periodic oscillation of $\theta$. 
This non-monotonic behavior means that $D^\mathrm{KL}(\hat{\rho}(\theta,0)||\hat{\rho}^{\rm SS}(\rho(\theta,0))$ is not the proper KL divergence. 

\begin{figure}
\centering
\includegraphics[clip,width=7cm]{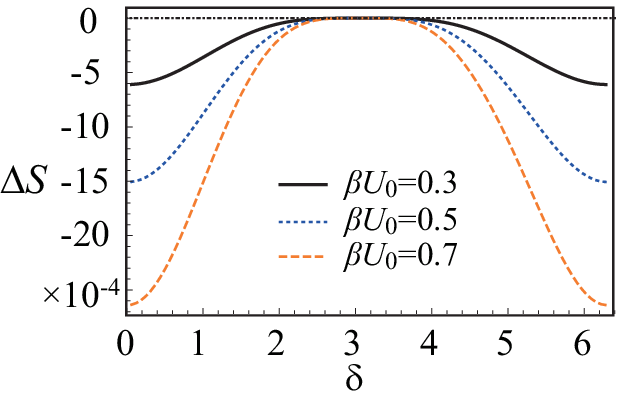}
\caption{Plots of $\Delta S(\delta)$ versus $\delta$ for $\beta U_0=0.3$ (solid line), $0.5$ (dotted line), $0.7$ (dashed line) with fixing $r=0.9$.
}
\label{fig4}
\end{figure}

Figure \ref{fig4} plots $\Delta S(\delta)$ versus $\delta$ for various $\beta U_0$ for $r=0.9$.
As can be seen in Eq.~\eqref{DeltaSDef} and Fig.~\ref{DKL}
the main contribution of $\Delta S(\delta)$ comes from the drastic change near $\theta=0$ where the mismatch between the geometric state and the initial steady state exists. 
We also note that $\Delta S(\delta)$ approaches zero in the limit $\beta U_0\to 0$.
This implies that the modulation of the Hamiltonian generates $\Delta S(\delta)$.
Although the periodic control of the Hamiltonian generates an electric current and the corresponding Joule heat 
even though the two reservoirs have no chemical potential difference, we have verified that their contribution is small.
Note that the expression of $\Delta S(\delta)$ does not have room to include such contributions. 

\begin{figure}
\centering
\includegraphics[clip,width=8cm]{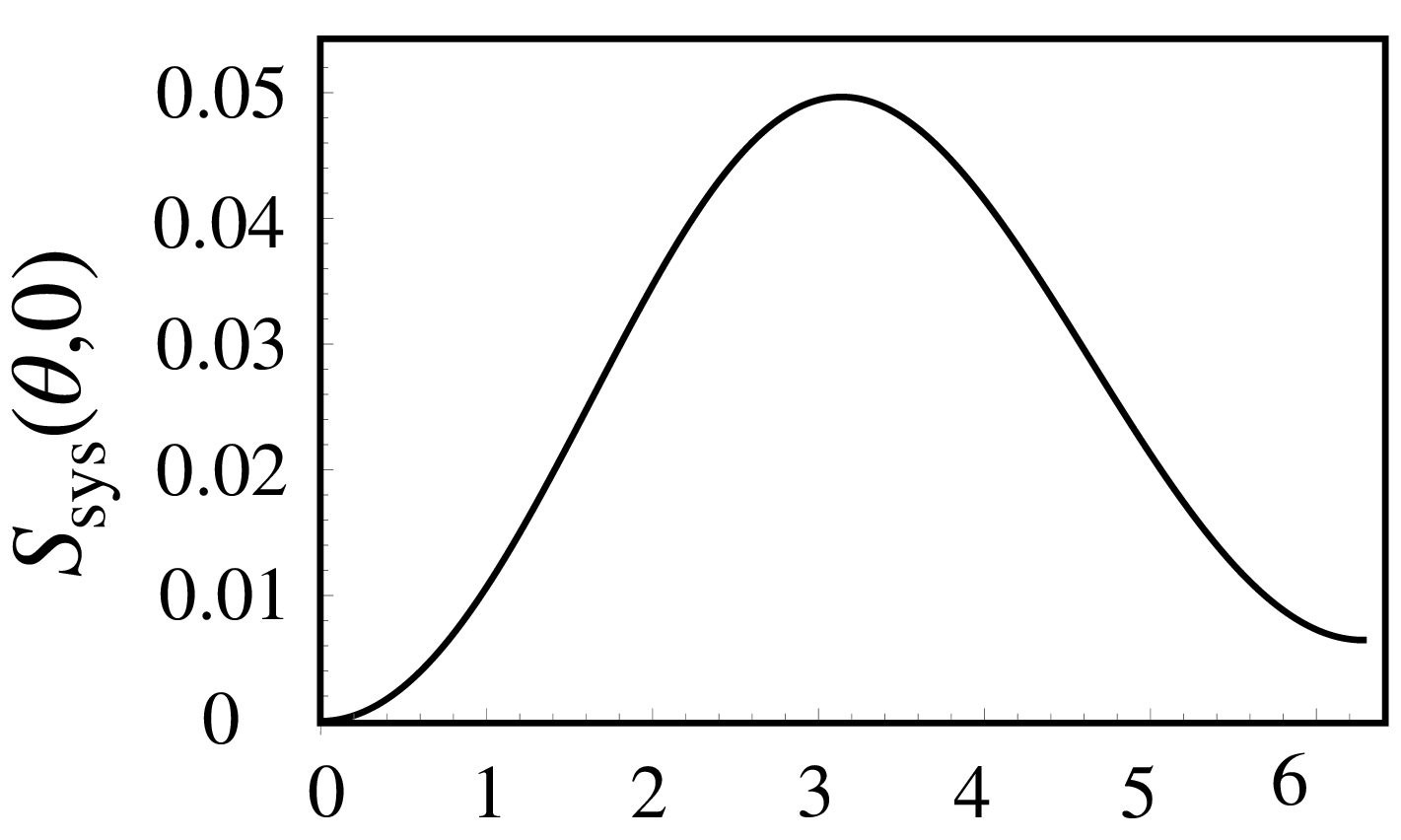}
\caption{Plot of $S_{\rm sys}(\theta,0)$ as a function of $\theta$ for $\beta U_0=0.1$ and $r=0.9$.  }
\label{tildeS}
\end{figure}

We also plot the system entropy $S_{\rm sys}(\theta,0)$ against $\theta$ in Fig.~\ref{tildeS}. 
As can be seen, $S_{\rm sys}(\theta,0)$ takes the maximum value at $\theta\simeq \pi$.
We also note that the behavior of $S^\mathrm{vN}(\theta,0)$ is presented in Appendix~\ref{App:Num_Th},
which is qualitatively different from $D^\mathrm{KL}(\hat{\rho}(\theta,0)||\hat{\rho}^\mathrm{GS}(\theta,0))$, $S_{\rm sys}(\theta,0)$ and ${\mathcal{Q}}(\theta,0)$.

\begin{figure}
\centering
\includegraphics[clip,width=8.6cm]{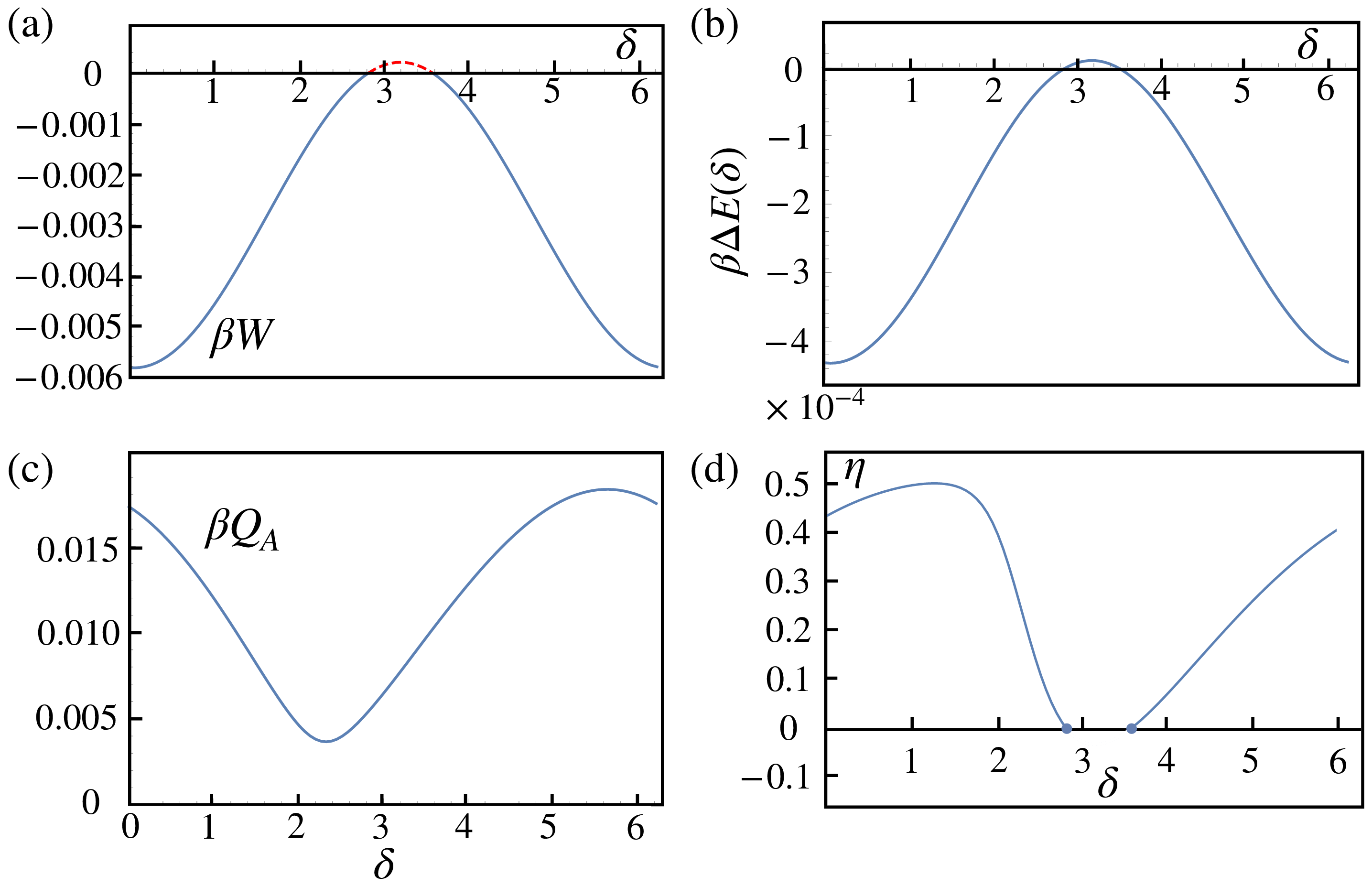}
\caption{
Plots of the work, the change of the internal energy, the absorbing heat, and the efficiency against $\delta$ for $\beta U_0=0.1$ and $r=0.9$. 
(a) Plot of $\beta W$ against $\delta$ (solid line). The region where the geometric demon is not achieved is shown as the dashed line. 
(b) Plot of $\beta \Delta E(\delta)$ against $\delta$ (solid line).
(c) Plot of $\beta \mathcal{Q}_A$ against $\delta$ (solid line).
(d) Plot of $\eta$ against $\delta$ (solid line). 
The region $2.824\le \delta\le 3.569$ is kept blank since the work becomes positive in the region.
} 
\label{fig6}
\end{figure}

We have confirmed that the work $W(\delta)$ introduced in Eq.\,\eqref{W} becomes negative except for the small parameter region around $\delta=\pi$ (i.e. $2.824\le \delta\le 3.569$), 
as shown in Fig.\,\ref{fig6}(a). 
This indicates that we can extract the work by cyclic modulations of the parameters in the Anderson model without fine-tuning in a wide range of parameters. 
The change of the energy in the first cycle $\Delta E(\delta)$ is shown in Fig.\,\ref{fig6}(b). 
We have verified the relationship $|\Delta E(\delta)|\ll |W(\delta)|$ and thus we have $|W(\delta)|\simeq |Q(\delta)|$ in this set of parameters. 
The heat absorbed during a modulation cycle is shown in Fig.\,\ref{fig6}(c).
Figure \ref{fig6}(d) shows the efficiency $\eta(\delta)$, defined in Eq.\,\eqref{efficiency}, for one modulation cycle. 
The efficiency is not defined for $2.824\le \delta\le 3.569$ because the work becomes positive in this region. 
These negative work and efficiency are the result of $\Delta S(\delta)<0$.
Thus, our engine is suitable for calling the geometric demon.


\section{Concluding Remarks}\label{sec:ConcludingRemarks}

We have implemented a geometric demon by modulating the electrochemical potentials in the two reservoirs and the repulsive $U(\theta)$ in the system Hamiltonian under isothermal conditions. 
We found that the state tends to relax towards the geometric state instead of the steady state when the parameter is kept modulated. 
The discrepancy between the steady state and the geometric state under a non-adiabatic modulation leads to the negative production of relative entropy. 
We can automatically extract the work from this engine with an increase in relative entropy if we start from the nonequilibrium steady state. 
We have verified that the work becomes negative in the wide range of the parameters for the Anderson model and thus the geometric demon can be implemented. 
Our geometric demon does not require any observation of states to reduce entropy. 
In this sense, our geometric demon can be easily implemented in realistic situations; 
therefore, we expect wide applications of this demon. 
Note that efficiency is unchanged even if we consider the second cycle. 
Indeed, we can stop the modulation after the first cycle and wait for the realization of the steady state, and we restart the modulation.

Our future tasks are as follows. 
(i) Since the present method of the argument is restricted to the case of $\epsilon\ll 1$, 
we will need to extend the analysis to the regime of larger $\epsilon$ as done in Ref.\,\cite{FHHT}. 
(ii) Although we have analyzed a pure quantum system to clarify entanglement or coherence effects~\cite{Brandner20,coherence,Um}, our treatment remains quasi-classical. 

\vspace*{0.5cm}
{\it Acknowledgements.-}
The authors thank Ville M. M. Paasonen, Kazutaka Takahashi, Kiyoshi Kanazawa, Sosuke Ito, Hikaru Watanabe, Asahi Yamaguchi and Fr\'{e}d\'{e}ric van Wijland for fruitful discussions and useful comments. 
This study was partially supported by a Grant-in-Aid of MEXT for Scientific Research KAKENHI (Grant Nos.\, 21H01006, 19K14616, and 20H01838).

\newpage

\appendix

\section{Time evolution of the density matrix}\label{appendix:initial}

This section consists of two subsections. 
In the first part, \ref{general_initial}, we derive the time evolution of $\hat{\rho}(\theta,\delta)$ 
from the general initial condition to demonstrate the universality of Eqs.\,(\ref{BSNConnection}) and (\ref{rhoCoeff}) for $\theta\gg \epsilon$.
In the second part, \ref{BSN_Suppl}, we present the detailed derivations of the BSN connection and BSN curvature.

\subsection{Derivation of the time-dependent expression of the density matrix starting from the general initial condition}\label{general_initial}

\subsubsection{Time evolution and geometrical state}

The purpose of this section is to derive Eqs.\,(\ref{BSNConnection}) and (\ref{rhoCoeff}).
Although we assume in the main text that the initial state is given by $\hat{\rho}^{\rm SS}(\theta=0,\delta)$, 
we can still derive these equations even when if we start with the generalized initial condition:
\begin{equation}
|\hat \rho_{\mathrm{ini}}\rangle=\sum_i a_i|r_i(0,\delta)\rangle, 
\label{geninirho}
\end{equation}
where $a_i$ is given by
\begin{equation}
a_i:=\langle \ell_i|\hat \rho_{\mathrm{ini}}\rangle.
\label{geniniai}
\end{equation}
The normalization of the density matrix leads to the condition: 
\begin{equation}
\mathrm{Tr}\hat\rho_{\mathrm{ini}}=\langle\ell_0|\hat \rho_{\mathrm{ini}}\rangle=\sum_i a_i\langle\ell_0|r_i(0,\delta)\rangle=a_0=1. 
\end{equation}

The time evolution of $|\rho(\theta,\delta)\rangle$ under Eq.\,\eqref{master} is given formally as 
\begin{align}
&|\hat{\rho}(\theta,\delta)\rangle
=\mathcal{T}\exp\left( \int^\theta_0 d\phi \hat K(\phi,\delta)\right)|\hat \rho_{\mathrm{ini}}\rangle,
\label{rhoformal}
\end{align}
As shown in the main text, the formal expression Eq.~\eqref{rhoformal} quickly approaches the geometric state $|\hat{\rho}^{\rm{GS}}(\theta,\delta)\rangle$. 
Although the geometric state is defined in the limit $\theta\to \infty$ in the strict sense, 
the difference between Eq.\,(\ref{rhoformal}) and the geometric state $|\hat{\rho}^{\rm{GS}}(\theta,\delta)\rangle$ decreases exponentially as $\theta$ increases. 
We note that Eq.\,\eqref{rhoformal} satisfies the normalization condition
\begin{align}
&\mathrm{Tr}\hat{\rho}(\theta,\delta)=
\langle \ell_0|\hat{\rho}(\theta,\delta)\rangle=\langle \ell_0|\hat \rho_{\mathrm{ini}}\rangle=1 ,
\label{rhoformalnormalization}
\end{align}
because $\langle \ell_0|$ is the left zero-eigenstate of $\hat K(\phi,\delta)$. 
Before entering the detailed calculation, we briefly explain the late time behavior of Eq.\,\eqref{rhoformal}. 
As we will see below, except for $i=0$ the term in Eq.\,\eqref{geninirho} is exponentially damped, and since $a_0=1$ the late time behavior hardly depends on the choice of the initial state. 

\subsubsection{Path-integral representation of the state}

Let us use the pass-integral-like expression for $|\hat{\rho}(\theta,\delta)\rangle$ as
\begin{align}
&|\hat{\rho}(\theta+\Delta \theta,\delta)\rangle\notag\\
&=e^{\epsilon^{-1}\hat K(\theta) \Delta \theta} |\hat{\rho}(\theta,\delta)\rangle\notag\\
&=e^{\epsilon^{-1}\hat K(\theta)\Delta \theta}\left( \sum_m |r_m(\theta,\delta)\rangle\langle \ell_m(\theta,\delta)| \right)|\hat{\rho}(\theta,\delta)\rangle\notag\\
&=\sum_m |r_m(\theta,\delta)\rangle e^{\varepsilon_m(\theta)\frac{\Delta \theta}{\epsilon}}\langle \ell_m(\theta,\delta)|\hat{\rho}(\theta,\delta)\rangle.
\label{rhotimeevolution}
\end{align}
Repeating Eq.~\eqref{rhotimeevolution} we can rewrite $|\hat{\rho}(\theta+\Delta \theta,\delta)\rangle$ as 
\begin{align}
&|\hat{\rho}(\theta+\Delta \theta,\delta)\rangle\notag\\
&=\sum_{m,n} |r_m(\theta,\delta)\rangle e^{\varepsilon_m(\theta)\frac{\Delta \theta}{\epsilon}}
\langle \ell_m(\theta,\delta) |r_m(\theta-\Delta \theta,\delta)\rangle \notag\\
&\hspace{3em}e^{\varepsilon_n(\theta-\Delta \theta)\frac{\Delta \theta}{\epsilon}}
\langle \ell_m(\theta-\Delta\theta,\delta)|\hat{\rho}(\theta-\Delta \theta,\delta)\rangle.
\label{rhotimeevolution2}
\end{align}
Thus, the recursive calculation yields 
\begin{align}
&|\hat{\rho}(\theta+\Delta \theta,\delta)\rangle\notag\\
&=  \sum_{i,j,k,\cdots,l} |r_i(\theta,\delta)\rangle e^{\varepsilon_i(\theta,\delta)\frac{\Delta \theta}{\epsilon}}\langle \ell_i(\theta,\delta)|r_j(\theta-\Delta \theta,\delta)\rangle\notag\\
&\hspace{3em}\times e^{\varepsilon_j(\theta-\Delta\theta,\delta)\frac{\Delta \theta}{\epsilon}}\langle \ell_j(\theta-\Delta \theta,\delta)|r_k(\theta-2\Delta\theta,\delta)\rangle\notag\\
&\hspace{3em}\cdots \times e^{\varepsilon_l(\Delta\theta)\frac{\Delta \theta}{\epsilon}}\langle \ell_l(\Delta\theta,\delta) |\hat{\rho}(0,\delta)\rangle\notag\\
&=  \sum_{i,j,k,\cdots,l,m}  a_n|r_i(\theta,\delta)\rangle e^{\varepsilon_i(\theta,\delta)\frac{\Delta \theta}{\epsilon}}\langle \ell_i(\theta,\delta)|r_j(\theta-\Delta \theta,\delta)\rangle\notag\\
&\hspace{3em}\times e^{\varepsilon_j(\theta-\Delta\theta,\delta)\frac{\Delta \theta}{\epsilon}}\langle \ell_j(\theta-\Delta \theta,\delta)|r_k(\theta-2\Delta\theta,\delta)\rangle\notag\\
&\hspace{3em}\cdots \times e^{\varepsilon_l(\Delta\theta)\frac{\Delta \theta}{\epsilon}}\langle \ell_l(\Delta\theta,\delta)|r_n(0,\delta)\rangle, 
\label{rhotimeevolution3}
\end{align}
where we have adopted the general initial condition $|\hat{\rho}(0,\delta)\rangle=|\hat \rho_{\mathrm{ini}}\rangle=\sum_n a_n|r_n(0,\delta)\rangle$. 

\begin{figure}
\centering
\includegraphics[clip,width=8.5cm]{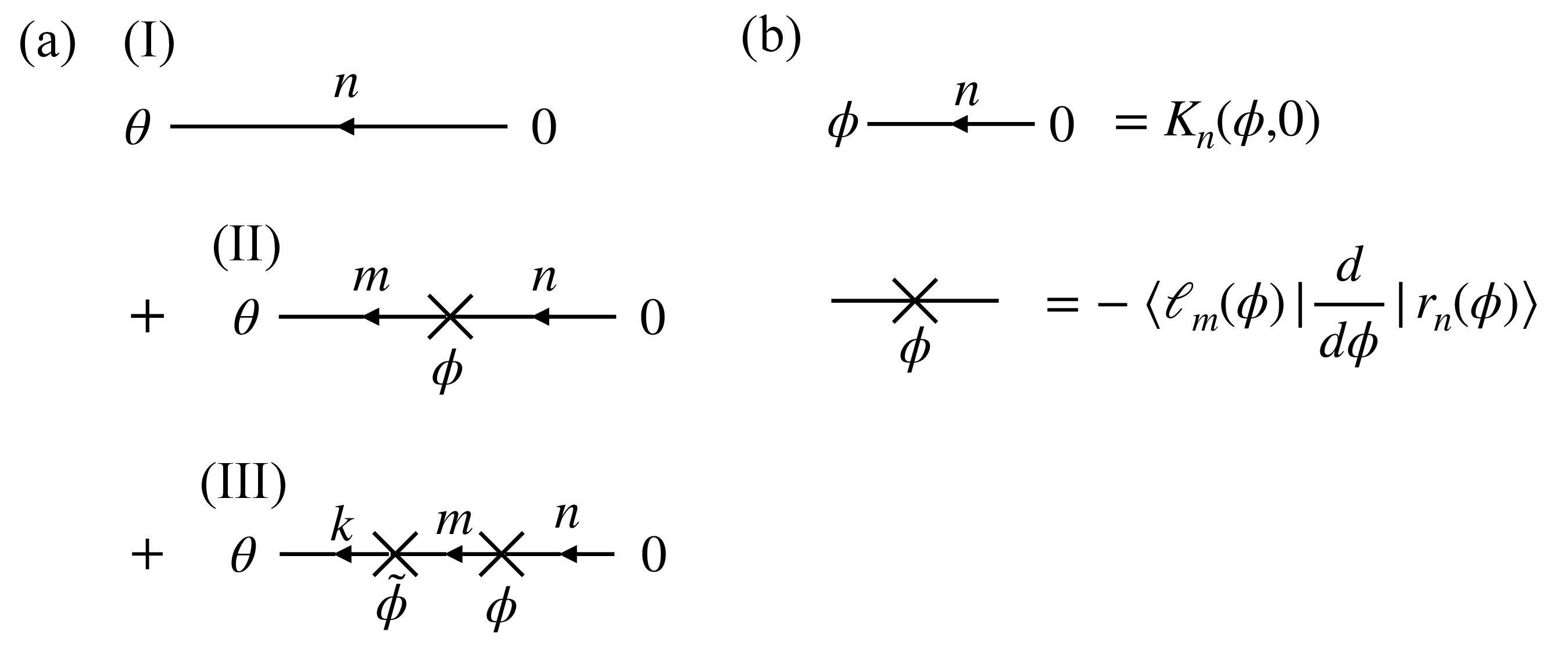}
\caption{Diagrammatic expressions of the Born type expansion. 
Figure \ref{diagram}\,(a) shows the 0th order (I), 1st order (II), and 2nd order (III) processes, respectively. 
The state evolves with $\theta$ by the free relaxation process (straight line) until the scattering process (represented by the cross mark) takes place (Fig. \ref{diagram}\,(b)). }
\label{diagram}
\end{figure}

Next we evaluate the connection (transfer matrix) $\langle \ell_i(\theta,\delta)|r_j(\theta-\Delta \theta,\delta)\rangle$ that connects the states in the different $\theta$. 
The connection can be calculated as follows up to the first order of $\Delta \theta$. 
\begin{align}
&\langle \ell_i(\theta,\delta)|r_j(\theta-\Delta \theta,\delta)\rangle\notag\\
&=\langle \ell_i(\theta,\delta)|\left(|r_j(\theta,\delta)\rangle-\Delta \theta \frac{d}{d\theta}|r_j(\theta,\delta)\rangle\right)\notag\\
&=\delta_{ij}-\Delta\theta \langle \ell_i(\theta,\delta)|\frac{d}{d\theta}|r_j(\theta,\delta)\rangle.
\label{connection}
\end{align}
The two terms in the last expression of Eq.\,\eqref{connection} have a natural interpretation; 
the first describing the second free propagation, and the second describing the jump process from the state $j$ to the state $i$. 
If we substitute Eq.\,\eqref{connection} into Eq.\,\eqref{rhotimeevolution3}, we obtain the expression: 
\begin{align}\label{Born-likeExpansion}
&|\hat{\rho}(\theta,\delta)\rangle=|\hat{\rho}(\theta,\delta)\rangle_0+|\hat{\rho}(\theta,\delta)\rangle_1+|\hat{\rho}(\theta,\delta)\rangle_2+\cdots.  
\end{align}
The first term on the RHS of Eq.\,(\ref{Born-likeExpansion}) is the free relaxation:
\begin{align}
|\hat{\rho}(\theta,\delta)\rangle_0
&=\sum_{n} a_n|r_n(\theta,\delta)\rangle K_n(\theta,0),
\end{align}
where we have introduced $K_n(b,a)$ as 
\begin{align}
K_n(b,a):=e^{\int_a^b \varepsilon_n(\phi,\delta)\frac{d\phi}{\epsilon}}. 
\end{align}
The second term on the RHS of Eq.~\eqref{Born-likeExpansion} contains a jump process:
\begin{align}
&|\hat{\rho}(\theta,\delta)\rangle_1
=\sum_{m,n} C_{mn}(\theta,\delta)a_n|r_m(\theta,\delta)\rangle, 
\end{align}
where the coefficient $C_{mn}$ is given by 
\begin{align}
C_{mn}(\theta,\delta):= \int_0^\theta d\phi K_m(\theta,\phi)J_{mn}(\phi) K_n(\phi,0)
\label{coeffgenini}
\end{align}
with the jump operator 
\begin{align}
J_{mn}(\phi):=-\langle \ell_m(\phi,\delta)|\frac{d}{d\phi}|r_n(\phi,\delta)\rangle.
\end{align}
The first and the second terms on the RHS of Eq.\,\eqref{Born-likeExpansion} can be understood by 
the diagrams shown in Figs.\,\ref{diagram}(a) (I) and (II), 
where the straight line represents the free relaxation and the cross mark represents the jump process. 
The first term on the RHS of Eq.\,\eqref{Born-likeExpansion} represents the free relaxation without any jump process. 
The second term on the RHS of Eq.\,\eqref{Born-likeExpansion}, 
the initial state $\sum_n a_n|r_n(0)\rangle$ evolves with $\theta$ by $K_n(\phi,0)$ and then the state $|r_n\rangle$ jumps to the state 
$|r_m\rangle$ at $\phi$ and then it evolves with $\theta$ by $K_m(\theta,\phi)$.  
Using Fig.\,\ref{diagram}(a)(III), we can write the third term on the RHS of Eq.\,\eqref{Born-likeExpansion} 
\begin{align}\label{secondBorn}
|\hat{\rho}(\theta,\delta)\rangle_2=\sum_{m,n,k} C_{kmn}(\theta)a_n|r_k(\theta,\delta)\rangle,
\end{align}
where
\begin{align}\label{C_kmn}
&C_{kmn}(\theta,\delta)\notag\\
&= \int_0^\theta d\phi \int_\phi^\theta d\phi^\prime  K_k(\theta,\phi^\prime)J_{km}(\phi^\prime)\notag\\
&\hspace{7em} \times K_m(\phi^\prime,\phi) J_{mn}(\phi) K_n(\phi,0). 
\end{align}

\subsubsection{Born-like approximation and further simplification}

In the following, we assume that the modulation speed is slow enough to ignore higher-order jump processes.
Thus, we consider only the first and the second terms on the RHS of Eq.\,\eqref{Born-likeExpansion}. 
This approximation is analogous to the Born approximation in the scattering theory. 
As emphasized in the main text, the state vector quickly approaches the geometric state with the Born-like approximation which is independent of the initial condition but depends on the BSN connection term. 

In the Born-like approximation, the resulting state is  
\begin{align}
&|\hat{\rho}(\theta,\delta)\rangle_0+|\hat{\rho}(\theta,\delta)\rangle_1\notag\\
&=\sum_m a_me^{\int_0^\theta\varepsilon_m(\phi,\delta)\frac{d\phi}{\epsilon}}|r_m(\theta,\delta)\rangle+\sum_{m,n}C_{mn}a_n|r_m(\theta,\delta)\rangle. 
\label{rhogenini}
\end{align}
The phase factor in the first term on RHS of Eq.\,(\ref{rhogenini}) is the dynamical phase, while the second term on the RHS is the term generated by the geometric phase. 

Since $a_0=1$ and $\varepsilon_0=0$ 
we can write Eq.\,\eqref{rhogenini} as 
\begin{align}
&|\hat{\rho}(\theta,\delta)\rangle_0+|\hat{\rho}(\theta,\delta)\rangle_1\notag\\
&= |r_0(\theta,\delta)\rangle+\sum_{m\neq 0} a_me^{\int_0^\theta\varepsilon_m(\phi,\delta)\frac{d\phi}{\epsilon}}|r_m(\theta,\delta)\rangle\notag\\
&\quad +\sum_{m}C_{m0}|r_m(\theta,\delta)\rangle+\sum_{m\neq 0,n\neq 0}C_{mn}a_n|r_m(\theta,\delta)\rangle\notag\\
&\quad +\sum_{n}C_{0n}a_n|r_m(\theta,\delta)\rangle. 
\label{rhogenini3}
\end{align}
The coefficient in the last term on the RHS of Eq.\,\ref{rhogenini3} is zero as
\begin{align}\label{A20}
C_{0n}=&-\int^\theta_0d\phi e^{\int_0^\phi\varepsilon_n(\zeta,\delta)\frac{d\zeta}{\epsilon}}\langle \ell_0(\phi, \delta)|\frac{d}{d\phi}|r_n(\phi, \delta)\rangle
\notag\\
=&-e^{\int_0^\phi\varepsilon_n(\zeta,\delta)\frac{d\zeta}{\epsilon}}
\left\{ \langle \ell_0|r_n(\theta, \delta)\rangle- \langle \ell_0|r_n(0, \delta)\rangle\right\}
\notag\\
&+\int^\theta_0d\phi \frac{d}{d\phi}\left(e^{\int_0^\phi\varepsilon_n(\zeta,\delta)\frac{d\zeta}{\epsilon}}\langle \ell_0|\right)|r_n(\phi, \delta)\rangle\notag\\
=&\delta_{0n}\int^\theta_0d\phi \frac{d}{d\phi}\left(e^{\int_0^\phi\varepsilon_n(\zeta,\delta)\frac{d\zeta}{\epsilon}}\right)=0, 
\end{align}
where we have used that $\langle\ell_0(\phi,\delta)|$ is independent of $\phi$ and $\delta$, and thus $\frac{d}{d\phi}\langle\ell_0(\phi,\delta)|=0$. 
We have also used the orthonormal relation $\langle\ell_m|r_n\rangle=\delta_{mn}$. 
The last equality in Eq.~\eqref{A20} holds since $\delta_{0n}=0$ for $n\neq 0$ and $\frac{d}{d\phi}\left(e^{\int_0^\phi\varepsilon_n(\zeta,\delta)\frac{d\zeta}{\epsilon}}\right)=0$ for $n=0$. 
Thus, Eq.\,\eqref{rhogenini3} becomes
\begin{align}
&|\hat{\rho}(\theta,\delta)\rangle_0+|\hat{\rho}(\theta,\delta)\rangle_1\notag\\
&= |r_0(\theta,\delta)\rangle+\sum_{m\neq 0} \left[a_me^{\int_0^\theta\varepsilon_m(\phi,\delta)\frac{d\phi}{\epsilon}}+C_{m0}\right]|r_m(\theta,\delta)\rangle\notag\\
&\quad +\sum_{m\neq 0,n\neq 0}C_{mn}a_n|r_m(\theta,\delta)\rangle 
\label{rhogenini4}
\end{align}
Note that the coefficient $C_{m0}$ is equivalent to $C_m$ in the main text.
We also note that we can directly obtain the relation
$\langle \ell_0|\hat \rho(\theta,\delta)\rangle_{n\ge 1}=0$ from trace preservation

\subsubsection{Late time behavior}
Now, let us consider the behavior for $\theta/\epsilon\gg 1$. 
It is obvious that the second term on the RHS of Eq.\,\eqref{rhogenini4} is negligible, because of the exponential damping factor that appears for $m\ne 0$. 
We also note that the sum of the first and third terms on the RHS is equivalent to $|\hat{\rho}(\theta,\delta)\rangle$ used in Eq.~\eqref{BSNConnection}.
Thus, Eq.\,\eqref{rhogenini4} is reduced to
\begin{align}
&|\hat{\rho}(\theta,\delta)\rangle_0+|\hat{\rho}(\theta,\delta)\rangle_1\notag\\
&= |\hat{\rho}^{\rm GS}(\theta,\delta)\rangle+\sum_{m,n}C_{mn}(\theta,\delta)a_n|r_m(\theta,\delta)\rangle
\label{relaxation}
\end{align}
for $\theta/\epsilon \gg 1$. 
It is straightforward to evaluate $C_{mn}(\theta,\delta)$ for $n\ne 0$ as
\begin{align}
|C_{mn}(\theta)|
=&\left|\int_0^{\theta}d\phi  e^{\int_\phi^\theta \varepsilon_m(\xi)\frac{d\xi}{\epsilon}}e^{\int_0^\phi \varepsilon_n(\zeta)\frac{d\zeta}{\epsilon}}\langle\ell_m(\phi) |\frac{d}{d\phi}|r_n(\phi)\rangle\right|\nonumber\\
\le&\int_0^{\theta}d\phi  e^{\int_\phi^\theta \varepsilon_m(\xi)\frac{d\xi}{\epsilon}}e^{\int_0^\phi \varepsilon_n(\zeta)\frac{d\zeta}{\epsilon}}\left|\langle\ell_m(\phi) |\frac{d}{d\phi}|r_n(\phi)\rangle\right|\nonumber \\
\le&\int_0^{\theta}d\phi  e^{\int_0^\theta \mathrm{max}(\varepsilon_m,\varepsilon_n)(\xi)\frac{d\xi}{\epsilon}}\left|\langle\ell_m(\phi) |\frac{d}{d\phi}|r_n(\phi)\rangle\right|,
\label{exponential_factor}
\end{align}
where we have omitted the $\delta$ dependence in Eq.\,\eqref{exponential_factor}.
Thus, $C_{mn}$ with $n\ne 0$ is much smaller than $C_{m0}$ because of the existence of the exponential factor in Eq.\,\eqref{exponential_factor}.
It can be shown that this late time behavior does not depend on the initial state since the coefficient $C_{m0}(\theta,\delta)$ depends only on the steady state $|r_0(\phi,\delta)\rangle$ and the states $|r_m(\phi,\delta)\rangle$ for $|\phi-\theta|\ll \theta_0$, where $\theta_0$ is the largest relaxation time of the states $|r_m(\phi,\delta)\rangle$. 
It is easy to show that the higher Born corrections are also independent of the initial state. 
Thus the geometric state can be approximated as 
\begin{align}
&|\hat{\rho}^{\rm GS}(\theta,\delta)\rangle\notag\\
&\simeq |\hat{\rho}^{\rm SS}(\theta,\delta)\rangle+\sum_{m}C_{m0}(\theta,\delta)|r_m(\theta,\delta)\rangle\ (\theta\gg \theta_0), 
\label{geometricalstateapprox}
\end{align}
where the equality holds up to the negligibly small deviation as in the case of Eq.\,(\ref{GS}). 
Here we note that \eqref{geometricalstateapprox} hardly depends on the choice of the initial state since 
\begin{align}\label{A20}
C_{m0}=&-\int^\theta_0d\phi e^{\int_\phi^\theta\varepsilon_m(\zeta,\delta)\frac{d\zeta}{\epsilon}}\langle \ell_m(\phi, \delta)|\frac{d}{d\phi}|r_0(\phi, \delta)\rangle
\notag\\
\simeq &-\int^\theta_{\theta_0}d\phi e^{\int_\phi^\theta\varepsilon_m(\zeta,\delta)\frac{d\zeta}{\epsilon}}\langle \ell_m(\phi, \delta)|\frac{d}{d\phi}|r_0(\phi, \delta)\rangle, 
\end{align}
where the difference between the RHS of the first line and that of the second line is negligibly small due to the exponential damping factor.

\subsection{Derivation of BSN connection and BSN curvature}\label{BSN_Suppl}

For the cyclic modulation $|r_i(2\pi,\delta)\rangle=|r_i(0,\delta)\rangle$, 
the time evolution from the general initial state becomes 
\begin{align}
\Delta |\hat{\rho}\rangle
=&\sum_{i\neq 0} a_i \left(e^{\int_0^{2\pi}\varepsilon_i(\phi,\delta)\frac{d\phi}{\epsilon}}-1\right)|r_i(0,\delta)\rangle\notag\\
&+\sum_{i,j}\mathcal{C}_{ij}a_je^{\int_0^{2\pi}\varepsilon_i(\phi,\delta)\frac{d\phi}{\epsilon}}|r_i (0,\delta)\rangle,\label{BSNConnectionGenIni}
\end{align}
where
\begin{align}
\mathcal{C}_{ij}:=&\delta_{ij}\oint_{\partial \Omega} d\phi \frac{\varepsilon_i(\phi,\delta)}{\epsilon}-\oint_{\partial \Omega} d \Lambda_\mu\mathcal{A}^\mu_{ij} 
\label{BSNCoeffGenIni}
\end{align} 
with the introduction of the BSN connection $\mathcal{A}^\mu_{ij}$:
\begin{align}
\mathcal{A}^\mu_{ij}:=
\langle\tilde{\ell}_i(\phi,\delta) |\frac{\partial}{\partial \Lambda_\mu}|\tilde{r}_j(\phi,\delta)\rangle,
\label{DefBSNConnectionGenIni}
\end{align} 
where we have introduced the gauged vectors 
\begin{align}
&\langle\tilde{\ell}_i(\phi,\delta) |:=\langle \ell_i(\phi,\delta) | e^{-\epsilon^{-1}\int_0^\phi\varepsilon_i(\zeta,\delta) \frac{d\zeta}{\epsilon}},\\
& |\tilde{r}_j(\phi,\delta)\rangle:=e^{\int_0^\phi\varepsilon_m(\zeta,\delta)\frac{d\zeta}{\epsilon}}|\tilde{r}_j(\phi,\delta)\rangle. 
\end{align} 
Here, the sum of the first term on the RHS of Eq.\,\eqref{BSNConnectionGenIni} is taken except for $i=0$ 
because $\varepsilon_0(\phi,\delta)=0$ always gives $e^{\int_0^\theta \varepsilon_0(\phi,\delta)\frac{d\phi}{\epsilon}}=1$. 
Using the Stokes theorem, we can rewrite Eq.\,(\ref{BSNCoeffGenIni}) as 
\begin{align}
\mathcal{C}_{ij}=\delta_{ij}\oint_{\partial \Omega} d\phi \frac{\varepsilon_i(\phi,\delta)}{\epsilon} - \int_{\Omega} d S_{\mu\nu} F^{\mu\nu}_{ij},  
\label{BSNCoeffGenIni2}
\end{align}  
where $\Omega$ is the area enclosed by the closed trajectory $\partial \Omega$, 
$S_{\mu\nu}=\frac{1}{2}d\Lambda_\mu\wedge d \Lambda_\nu$, and $F^{\mu\nu}_{ij}$ represents the BSN curvatures defined as
\begin{align}
F^{\mu\nu}_{ij}:
=\frac{\partial \langle\tilde{\ell}_i|}{\partial \Lambda_\nu} \frac{\partial |\tilde{r}_j\rangle}{\partial\Lambda_\mu}
-\frac{\partial \langle \tilde{\ell}_i|}{\partial \Lambda_\mu} \frac{\partial |\tilde{r}_j\rangle}{\partial\Lambda_\nu}.
\label{DefBSNCurvaureGenIni}
\end{align}
It is also possible to rewrite Eq.\,(\ref{BSNCoeffGenIni2}) as 
\begin{align}
\mathcal{C}_{ij}&= \delta_{ij}\oint_{\partial \Omega} d\phi \frac{\varepsilon_i(\phi,\delta)}{\epsilon}+\frac{1}{2}\int_{\Omega} d \langle \tilde{\ell}_i| \wedge d|\tilde{r}_j\rangle.  
\label{BSNCoeffOCB3}
\end{align}

\section{Differences between the present study and the previous studies}\label{Diff_App}

This section is devoted to explaining the difference between the present study and the previous studies. 
In the first part, we show the consistency between the results in Ref.\,\cite{FHHT} and those in the present study. 
In the second part, we present the difference between the setup in the present study and the previous study \cite{Hayakawa2021}. 

\subsection{The relationship of our work with Ref.\,\cite{FHHT}}

Here, we briefly comment on the superficial differences between the present study and Ref.\,\cite{FHHT}. 
In the present analysis, we have fixed the gauge degrees of freedom to fix the overall scale of states. 
It is known that $\{|r_m(\theta)\rangle, \langle\ell_m(\theta)|\}$ and $\{e^{-f_m(\theta)}|r_m(\theta)\rangle, e^{f_m(\theta)}\langle\ell_m(\theta)|\}$ yield the same results. 
The gauge freedom can be fixed if we choose the scaling, e.g.\ $\langle \ell_0|=(1,1,1,1)$ for the Anderson model. 
We also note that the 0th-order Born approximation in this paper is called the adiabatic approximation in Ref.~\cite{FHHT}.

In Ref.\,\cite{FHHT}, the gauge degrees of freedom for the left eigenstates ($\langle \phi_n|$) and the right eigenstates ($|\phi_m\rangle$) are fixed by \footnote{We note that the lowest eigenstate is labeled by $m=1$ in Ref.\,\cite{FHHT} whereas it is labeled by $m=0$ in this paper. To avoid confusion, we relabel the subscripts $m$ for the eigenstates in Ref.\,\cite{FHHT} from $\{1,\cdots\}$ to $\{0,\cdots\}$}
\begin{align}
&|\tilde\phi_m (\theta)\rangle = e^{-\int^\theta_0d\eta\langle\phi_m(\eta)|\frac{d}{d\eta}|\phi_m(\eta)\rangle}|\phi_m (\theta)\rangle,\label{phiandtildephi}\\
&\langle\tilde\phi_m (\theta)| = e^{\int^\theta_0d\eta\langle\phi_m(\eta)|\frac{d}{d\eta}|\phi_m(\eta)\rangle}\langle\phi_m (\theta)|.\label{phiandtildephi2}
\end{align}
Since both states $|\phi_m\rangle, \langle \phi_m|$ and $|r_m\rangle,\langle \ell_m|$ are left and right eigenstates of $\hat K$ with eigenvalue $\varepsilon_m$, those are identical up to the scaling factor as 
\begin{align}
&|\phi_m(\theta)\rangle=e^{g(\theta)} |r_m(\theta)\rangle,\label{ellandphi}\\
&\langle \phi_m(\theta)|=e^{-g(\theta)}\langle \ell_m(\theta)|. \label{randphi}
\end{align}
Substituting Eqs.\,\eqref{ellandphi} and \eqref{randphi} into Eqs.\,\eqref{phiandtildephi} and \eqref{phiandtildephi2}, we obtain the relation between $\{|\tilde \phi_m(\theta)\rangle, \langle\tilde \phi_m(\theta)\}$ in Ref.\,\cite{FHHT} and 
$\{|r_m(\theta)\rangle, \langle\ell_m(\theta)\}$ in the present study as 
\begin{align}
&|\tilde \phi_m(\theta)\rangle=e^{g(0)+\int_0^\theta d\eta \langle \ell_m(\eta)|\frac{d}{d\eta}|r_m(\eta)\rangle } |r_m(\theta)\rangle,\label{prevandpresent}\\
&\langle \tilde\phi_m(\theta)|=e^{-g(0)-\int_0^\theta\langle \ell_m(\eta)|\frac{d}{d\eta}|r_m(\eta)\rangle }\langle \ell_m(\theta)|. \label{prevandpresent2}
\end{align}
By using $ \int_0^\theta d\eta\langle \ell_0(\eta)|\frac{d}{d\eta}|r_0(\eta)\rangle=\left[\langle \ell_0(\eta)|r_0(\eta)\rangle \right]^\theta_0-\int_0^\theta\frac{d\langle \ell_0(\eta)|}{d\eta}|r_0(\eta)\rangle=0$, one can show that the lowest eigenstate in Ref.\,\cite{FHHT} is identical to $|r_0\rangle, \langle\ell_0|$. 

We note that if we ignore the last term on the RHS of Eq.\,\eqref{rhogenini4}, our expression is reduced to Eq.\,(5) in Ref.\,\cite{FHHT}, 
where they assume that the time evolution does not make transitions to different eigenstates, which corresponds to the Born-type approximation in our approach. 
The other assumption is made in Ref.\,\cite{FHHT} which corresponds to setting $C_{mn}=0$ with $n\neq 0$ in Eq.\,\eqref{rhogenini4}. 
This approximation is also justified in our approach and will be explained in the next subsection. 
With these approximations, we can directly show the consistency between our results and those in Ref.\,\cite{FHHT} by substituting Eqs.\,\eqref{prevandpresent} and \eqref{prevandpresent2} into Eq.\,\eqref{rhogenini4}. 
In the main part of our paper, since we start from the steady state ($a_0=1, a_{m\neq0}$), $C_{mn}$ with $n\neq 0$ does not appear, and thus the latter assumption is not necessary,

\subsection{Difference between Ref.\,\cite{Hayakawa2021} and the present study}\label{app:Diff}

A major difference between this paper and Ref.\,\cite{Hayakawa2021} is the setup.
We mainly discuss the relaxation process from NESS in this paper, but Ref.\,\cite{Hayakawa2021} discussed the behavior after the system reaches a quasi-periodic state, in which the geometric term can be ignored.
Thus the previous study did not consider the initial relaxation and treated the steady state $\rho^{\rm SS}$.

It is obvious that $\hat{\rho}^\mathrm{GS}(\theta,\delta)$ is different from $\hat{\rho}^\mathrm{SS}(\theta,\delta)$
due to the existence of the BSN connection term as shown in Eqs.~\eqref{BSNConnection} and \eqref{rhoCoeff}.
Moreover, as shown in the main text, the mismatch between $\hat{\rho}^\mathrm{SS}(0,\delta)$ and $\hat{\rho}^\mathrm{GS}(\theta,\delta)$ for small $\theta$ plays a significant role.
Thus, if  we are interested in the initial relaxation process from $\hat{\rho}^\mathrm{SS}(\delta)$ to $\hat{\rho}^\mathrm{GS}(\theta,\delta)$,  
we cannot use the $\epsilon$ perturbation around $\hat{\rho}^\mathrm{SS}(\theta,\delta)$ used in Ref.\,\cite{Hayakawa2021}.

\section{Properties of the Anderson model}\label{app:Anderson}

This section summarizes the properties of the Anderson model in more detail, 
which consists of three subsections. 
The first subsection summarizes the evolution matrix.
In the second subsection, we provide the explicit forms of the BSN connection and BSN curvature as well as the expansion coefficients. 
In the third subsection, we present some detailed calculations for the Anderson model. 
For simplicity, we do not write the $\delta$ dependence of the variables explicitly in this section.

\subsection{Evolution matrix and eigenstates in the Anderson model}\label{appendix:eigen}

In this subsection, we summarize the evolution matrix and eigenstates in the Anderson model.
A similar discussion can be found in Ref.\,\cite{Hayakawa2021}.

Since $\hat{\rho}$ is a diagonal matrix, $|\hat{\rho}\rangle$ also has only four components, 
and the transition matrix $\hat{K}$ in Eq.\,(\ref{master}) in the wideband approximation is given by the $4\times 4$ matrix:
\begin{equation}\label{Ville_9_0806} 
\hat{K}=-\begin{pmatrix}
2f_-^{(1)} & -f_+^{(1)} & -f_+^{(1)} & 0 \\
-f_-^{(1)}& f_-^{(0)}+f_+^{(1)} & 0 & -f_+^{(0)} \\
-f_-^{(1)}& 0 & f_-^{(0)}+f_+^{(1)}  & -f_+^{(0)} \\
0 & -f_-^{(0)} & -f_-^{(0)} & 2f_+^{(0)} \\
\end{pmatrix} ,
\end{equation}
where we have introduced
\begin{align}
f_+^{(j)}:=f_\mathrm{L}^{(j)}+f_\mathrm{R}^{(j)},\ f_-^{(j)}:=2-f_+^{(j)},
\end{align}
with the Fermi distribution 
\begin{equation}\label{Fermi}
f_\alpha^{(j)}(\mu_\alpha(\theta),U(\theta)):=\frac{1}{1+e^{\beta (\epsilon_0+j U(\theta)-\mu_\alpha(\theta))}}
\end{equation}
in the lead $\alpha(=L\ {\rm or}\ R)$ for the single occupancy $j=0$ and double occupancy $j=1$. 
The eigenvalues and eigenvectors correspond to the evolution matrix $\hat{K}(\theta,\delta)$ in Eq.~\eqref{Ville_9_0806} are presented in Ref.~\cite{Hayakawa2021}.

\subsection{BSN connection for Anderson model}\label{appendix:BSNAnderson}

In this subsection, we present an explicit form of the BSN connection for the Anderson model.
Note that the calculation presented here is derived directly from the detailed expressions of the eigenfunctions and eigenvalues of Eq.~\eqref{master} (see Ref.~\cite{Hayakawa2021} for details).

For the explicit calculation of the BSN connection, we use Eqs.\,(\ref{BSNConnection}) and (\ref{rhoCoeff}) with $\theta=2\pi$. 
Integrating by parts, we get
\begin{align}
&\int_0^{2\pi}e^{\int_\phi^{2\pi}\varepsilon_2(\xi)\frac{d\xi}{\epsilon} }\langle  \ell_2(\phi)|\frac{d}{d\phi}|r_0(\phi)\rangle\notag\\
&=\int_0^{2\pi}e^{\int_\phi^{2\pi}\varepsilon_2(\xi)\frac{d\xi}{\epsilon}}
\left[\frac{d}{d\phi}\langle  \ell_2(\phi)|r_0(\phi)\rangle-\frac{d\langle  \ell_2(\phi)|}{d\phi}|r_0(\phi)\rangle\right]\notag\\
&=0,
\end{align}
where $\langle \ell_2(\phi)|$ is independent of $\phi$. 
Thus, the summation of $i$ in Eq.\,(\ref{BSNConnection}) is reduced to the summation with $i=1$ and 3. 
The differentiation of $|r_0(\phi)\rangle$ with respect to $\phi$ becomes 
\begin{align}\label{diffr0}
&\frac{d}{d\phi}|r_0(\phi)\rangle\notag\\
&=
\left[\frac{d}{d\phi}
\frac{1}{2(f_+^{(0)}+f_-^{(1)})}\right]
\begin{pmatrix}
f_+^{(0)}f_+^{(1)} \\[0.5em]  
f_+^{(0)} f_-^{(1)}\\[0.5em]   
f_+^{(0)} f_-^{(1)}\\[0.5em]  
f_-^{(0)} f_-^{(1)}
\end{pmatrix}\notag\\
&\hspace{1em}
+\frac{1}{2(f_+^{(0)}+f_-^{(1)})}
\frac{d}{d\phi}
\begin{pmatrix}
f_+^{(0)}f_+^{(1)} \\[0.5em]  
f_+^{(0)} f_-^{(1)}\\[0.5em]   
f_+^{(0)} f_-^{(1)}\\[0.5em]  
f_-^{(0)} f_-^{(1)}
\end{pmatrix}\notag\\
&=\alpha(\Lambda)|r_0\rangle+\frac{1}{2(f_+^{(0)}+f_-^{(1)})}
\frac{d}{d\phi}
\begin{pmatrix}
f_+^{(0)}f_+^{(1)} \\[0.5em]  
f_+^{(0)} f_-^{(1)}\\[0.5em]   
f_+^{(0)} f_-^{(1)}\\[0.5em]  
f_-^{(0)} f_-^{(1)}
\end{pmatrix},
\end{align}
where $\alpha(\Lambda)$ is an unimportant factor 
because we are only interested in $\langle \ell_i(\phi)|\frac{d}{d\phi}|r_0(\phi)\rangle$ when $i\ne 0$ and $\langle \ell_i(\phi)|r_0(\phi)\rangle=0$ for $i\ne 0$. 

Substituting Eq.\,(\ref{diffr0}) into Eq.\,(\ref{BSNConnection}) with $\theta=2\pi$ results in
\begin{align}\label{rhoexpl}
&|\hat{\rho}(2\pi)\rangle \simeq  |r_0(2\pi)\rangle+\sum_{1,3}\mathcal{C}_i|r_i (2\pi)\rangle,\\
&\mathcal{C}_i=-\int_0^{2\pi}d\phi \frac{\langle\tilde\ell_i(\phi) |}{2(f_+^{(0)}+f_-^{(1)})}\frac{d}{d\phi}
\begin{pmatrix}
f_+^{(0)}f_+^{(1)} \\[0.5em]  
f_+^{(0)} f_-^{(1)}\\[0.5em]   
f_+^{(0)} f_-^{(1)}\\[0.5em]  
f_-^{(0)} f_-^{(1)}
\end{pmatrix}. 
\end{align}
By using $\frac{d}{d\theta}(f_+^{(j)}+f_-^{(j)})=0$, $\mathcal{C}_i$ becomes 
\begin{align}\label{BSNCoeff}
\mathcal{C}_i=&-\int_0^{2\pi}d\phi \frac{\langle\tilde\ell_i(\phi) |}{2(f_+^{(0)}+f_-^{(1)})}\frac{df_+^{(0)}}{d\phi}
\begin{pmatrix}
f_+^{(1)} \\[0.5em]  
f_-^{(1)}\\[0.5em]   
f_-^{(1)}\\[0.5em]  
-f_-^{(1)}
\end{pmatrix}\notag\\
&-\int_0^{2\pi}d\phi \frac{\langle\tilde\ell_i(\phi) |}{2(f_+^{(0)}+f_-^{(1)})}\frac{df_+^{(1)}}{d\phi}
\begin{pmatrix}
f_+^{(0)} \\[0.5em]  
-f_+^{(0)}\\[0.5em]   
-f_+^{(0)}\\[0.5em]  
-f_-^{(0)} 
\end{pmatrix}. 
\end{align}
Substituting the expression of $\langle \ell_1|$ \cite{Hayakawa2021}
into Eq.\,(\ref{BSNCoeff}), we obtain 
\begin{align}
\mathcal{C}_1=&-\int_0^{2\pi}d\phi e^{\int_\phi^{2\pi}\varepsilon_1(\xi)\frac{d\xi}{\epsilon} }\frac{2f_-^{(1)}}{f_+^{(0)}+f_-^{(1)}}\frac{df_+^{(0)}}{d\phi}
\notag\\
&-\int_0^{2\pi}d\phi e^{\int_\phi^{2\pi}\varepsilon_1(\xi)\frac{d\xi}{\epsilon} }\frac{2f_+^{(0)}}{f_+^{(0)}+f_-^{(1)}}\frac{df_+^{(1)}}{d\phi}. 
\end{align}
Similarly, we obtain 
\begin{align}
\mathcal{C}_3=&\int_0^{2\pi}d\phi e^{\int_\phi^{2\pi}\varepsilon_3(\xi)\frac{d\xi}{\epsilon} }\frac{f_+^{(1)}f_-^{(1)}}{f_+^{(0)}+f_-^{(1)}}\frac{df_+^{(0)}}{d\phi}
\notag\\
&-\int_0^{2\pi}d\phi e^{\int_\phi^{2\pi}\varepsilon_3(\xi)\frac{d\xi}{\epsilon} }\frac{f_+^{(0)}f_-^{(0)}}{f_+^{(0)}+f_-^{(1)}}\frac{df_+^{(1)}}{d\phi}\notag\\
=&\int_0^{2\pi}d\phi e^{4\epsilon^{-1}(\phi-2\pi)}\frac{f_+^{(1)}f_-^{(1)}}{f_+^{(0)}+f_-^{(1)}}\frac{df_+^{(0)}}{d\phi}
\notag\\
&-\int_0^{2\pi}d\phi e^{4\epsilon^{-1}(\phi-2\pi)}\frac{f_+^{(0)}f_-^{(0)}}{f_+^{(0)}+f_-^{(1)}}\frac{df_+^{(1)}}{d\phi}. 
\label{coeffC3}
\end{align}

As shown by Eq.\,(\ref{coeffC3}), the factor $e^{4\epsilon^{-1}(\phi-2\pi)}$ in the integrand plays an important role. 
Because of this factor, it is not necessary to consider long-term memory in the dynamics.
In the case of $\mathcal{C}_3$, the scaling factor does not depend on the choice of the trajectory, but only depends on $\theta$.
Thus, one can estimate the BSN curvature at $\phi$.

For $\epsilon\ll1$, the exponential factor $e^{4\epsilon^{-1}(\phi-2\pi)}$ behaves as a cutoff function and thus 
\begin{align}
\mathcal{C}_3
\approx&\int_{2\pi-\epsilon}^{2\pi}d\phi
\left[\frac{f_+^{(1)}f_-^{(1)}}{f_+^{(0)}+f_-^{(1)}}\frac{df_+^{(0)}}{d\phi}-\frac{f_+^{(0)}f_-^{(0)}}{f_+^{(0)}+f_-^{(1)}}\frac{df_+^{(1)}}{d\phi}\right] \notag \\
\sim &\epsilon \left[\frac{f_+^{(1)}f_-^{(1)}}{f_+^{(0)}+f_-^{(1)}}\frac{df_+^{(0)}}{d\phi}-\frac{f_+^{(0)}f_-^{(0)}}{f_+^{(0)}+f_-^{(1)}}\frac{df_+^{(1)}}{d\phi}\right]_{\phi=2\pi}.&
\end{align}
Similarly, we can write 
\begin{align}
\mathcal{C}_1
\sim &
\epsilon\left[
\frac{2 e^{\varepsilon_1(2\pi)}}{f_+^{(0)}+f_-^{(1)}}\left(f_-^{(1)}\frac{df_+^{(0)}}{d\phi}-f_+^{(0)}\frac{df_+^{(1)}}{d\phi}\right)\right]_{\phi=2\pi}.
\end{align}
From the above estimates, the leading order contribution of the geometric phase is the order $O(\epsilon)$. 

Let us show that $\mathcal{C}_3$ reduces to zero in the non-interacting case $\beta U_0=0$. 
In this case, $f^{(0)}_\pm=f^{(1)}_\pm\equiv g_\pm$ and $g_++g_-=2$; thus $\mathcal{C}_{3}$ becomes
\begin{align}\label{BSNcoeffNoint}
\lim_{\beta U_0\to 0}\mathcal{C}_3&=0.
\end{align} 
In the opposite limit $\beta U_0\to \infty$ with $f_+^{(1)}=0,\ f_-^{(1)}=2$
$\mathcal{C}_{3}$ also becomes zero, because  
\begin{align}\label{BSNcoessUinf}
\lim_{\beta U_0\to \infty}\mathcal{C}_3&=0.
\end{align}

\subsection{Some detailed results for the Anderson model}\label{app:results}

In this subsection, we present some detailed results beyond the main text, as well as a figure showing the control parameters in the parameter space.

\begin{figure}
\centering
\includegraphics[clip,width=6cm]{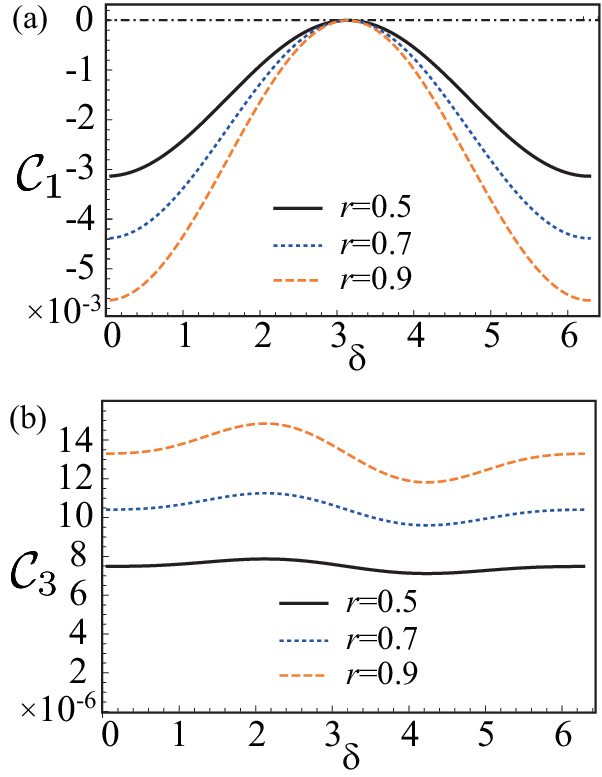}
\caption{Plots of $\mathcal{C}_1$ (Fig.\,(a)) and $\mathcal{C}_3$ (Fig.\,(b)) versus $\delta$ for $r=0.5,0.7,0.9$ with $\beta U_0=0.1$.
}
\label{fig1}
\end{figure}

In Fig.\,\ref{fig1}, we plot $\mathcal{C}_{1}(\delta)$ (left figure) and $\mathcal{C}_{3}(\delta)$ (right figure) versus $\delta$ for various $r$ and $\beta U_0=0.1$. 
As shown, the coefficients increase with increasing $r$. 
The coefficients $\mathcal{C}_1$ and $\mathcal{C}_3$ become zero at $\delta=\pi$. 

\begin{figure}
\centering
\includegraphics[clip,width=6cm]{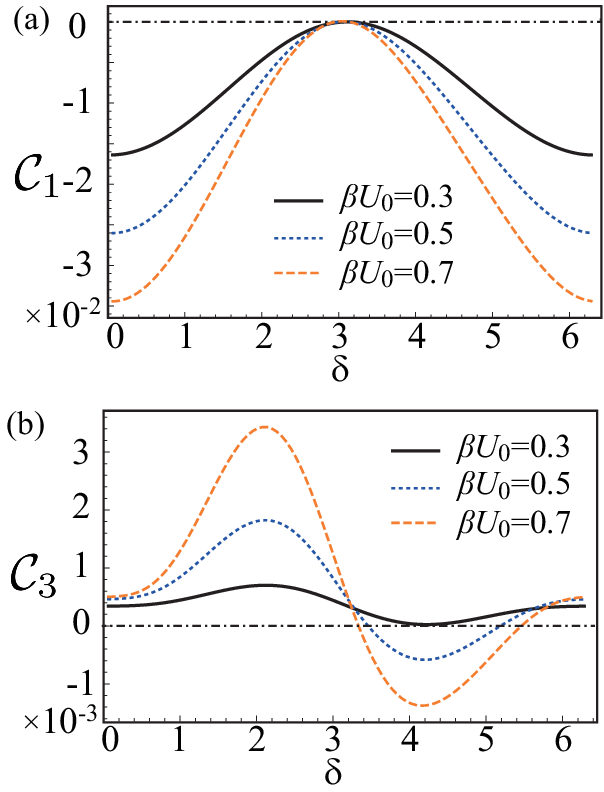}
\caption{Plots of $\mathcal{C}_1$ (Fig.\,(a)) and $\mathcal{C}_3$ (Fig.\,(b)) versus $\delta$ for $\beta U_0=0.3,\ 0.5,\ 0.7$ with $r=0.9$.
}
\label{fig2}
\end{figure}

Figure \ref{fig2} shows the coefficients $\mathcal{C}_{1}(\delta)$ (left figure) and $\mathcal{C}_{3}(\delta)$ (right figure) versus $\delta$ for various $\beta U_0$ and $r=0.9$. 
As shown in Fig.\,\ref{fig2}, $|\mathcal{C}_i|$ with $i=1$ and 3 increases with $\beta U_0$.


\section{Some detailed properties of the density matrix, von Neumann entropy and heat flow}\label{App:Num_Th}

In this section, we present the detailed results on the density matrix, the von Neumann entropy, and the heat flow.

\begin{figure}
\centering
\includegraphics[clip,width=6cm]{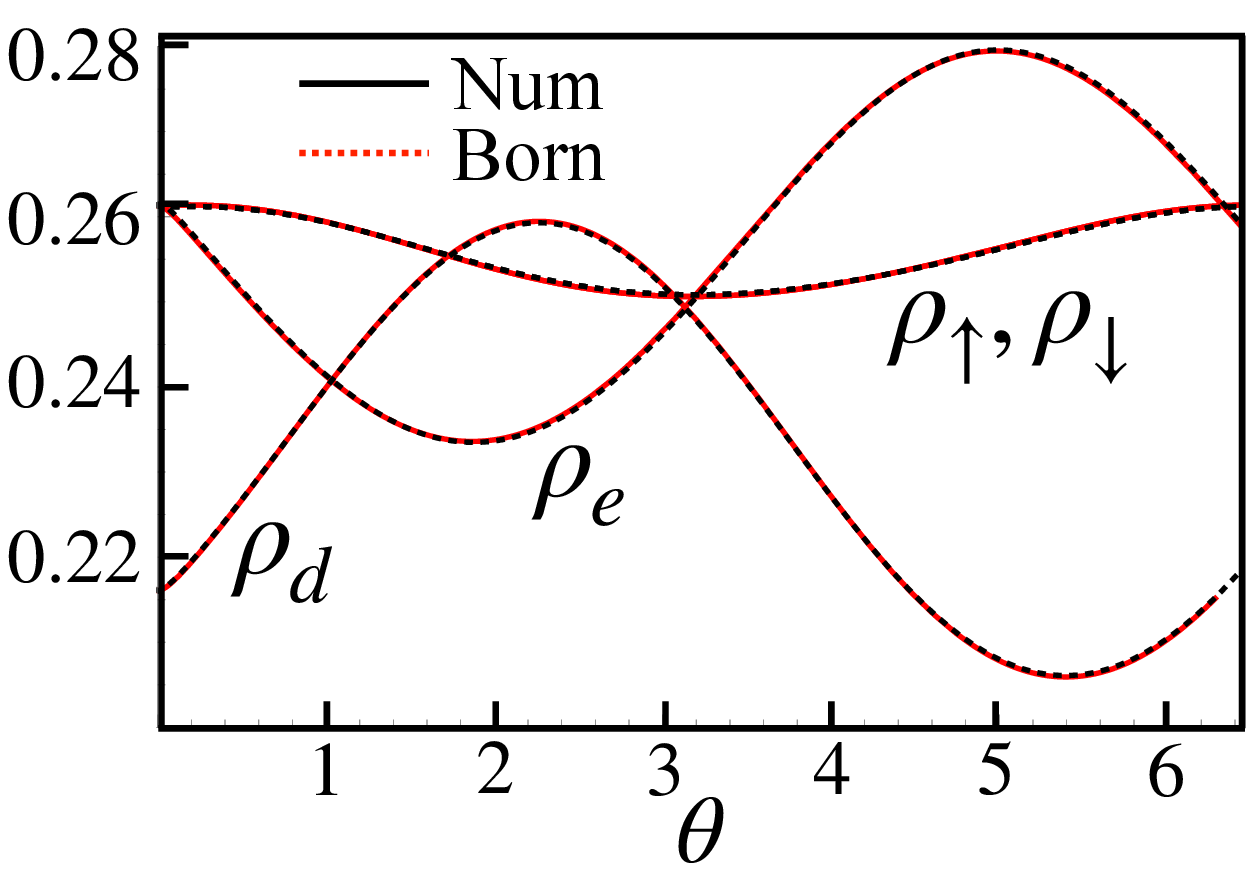}
\caption{Plots of the time evolution of the density matrix element based on the Born-like approximation (red-dashed line) and the numerical solutions of the master equation (black-solid line). 
The parameters are set to be $\beta \bar\mu=0.1$, $\beta U_0=0.1$, and $\beta \epsilon_0=0.1$.}
\label{rhoNum}
\end{figure}

First, we verify the validity of the Born-like approximation by comparing its solution with that obtained by the numerical solution of Eq.~\eqref{master}. 
In Fig.\,\ref{rhoNum}, we plot the time dependence of the diagonal elements of the density matrix computed by the Born-like approximation and the numerical solutions. 
This shows that the Born-like approximation gives us sufficiently accurate expressions since the deviation between the numerical solutions and the expressions based on the Born-like approximation is invisible.

\begin{figure}
\centering
\includegraphics[clip,width=7.5cm]{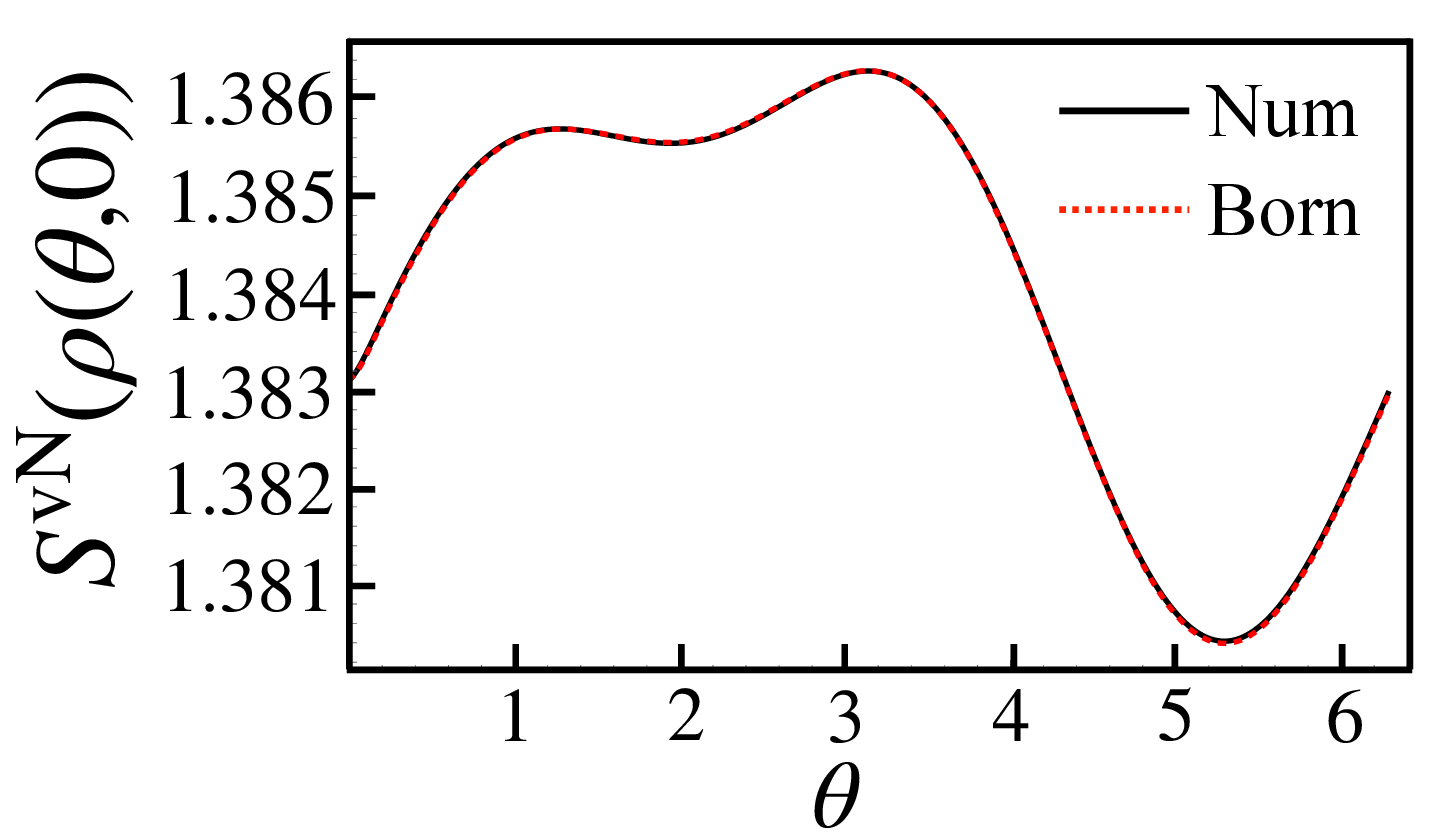}
\caption{Plot of von Neumann entropy as a function of $\theta$. 
The results obtained by the Born-like approximation (red-dashed line) and the numerical calculation of the master equation (black-solid line) are plotted.}
\label{vNNum}
\end{figure}

Next, we plot $S^\mathrm{vN}(\theta,0)$ against $\theta$ in Fig.~\ref{vNNum}.
This figure contains the comparison of $S^\mathrm{vN}(\theta,0)$ obtained by the Born-like approximation as in Eqs.~\eqref{BSNConnection} and \eqref{rhoCoeff} with that obtained by the numerical integration of Eq.~\eqref{master}.
As can be seen, the difference between the two results is invisible.
The behavior of $S^\mathrm{vN}(\theta,0)$ is neither non-monotonic nor quasi-periodic.

\begin{figure}
\centering
\includegraphics[clip,width=7cm]{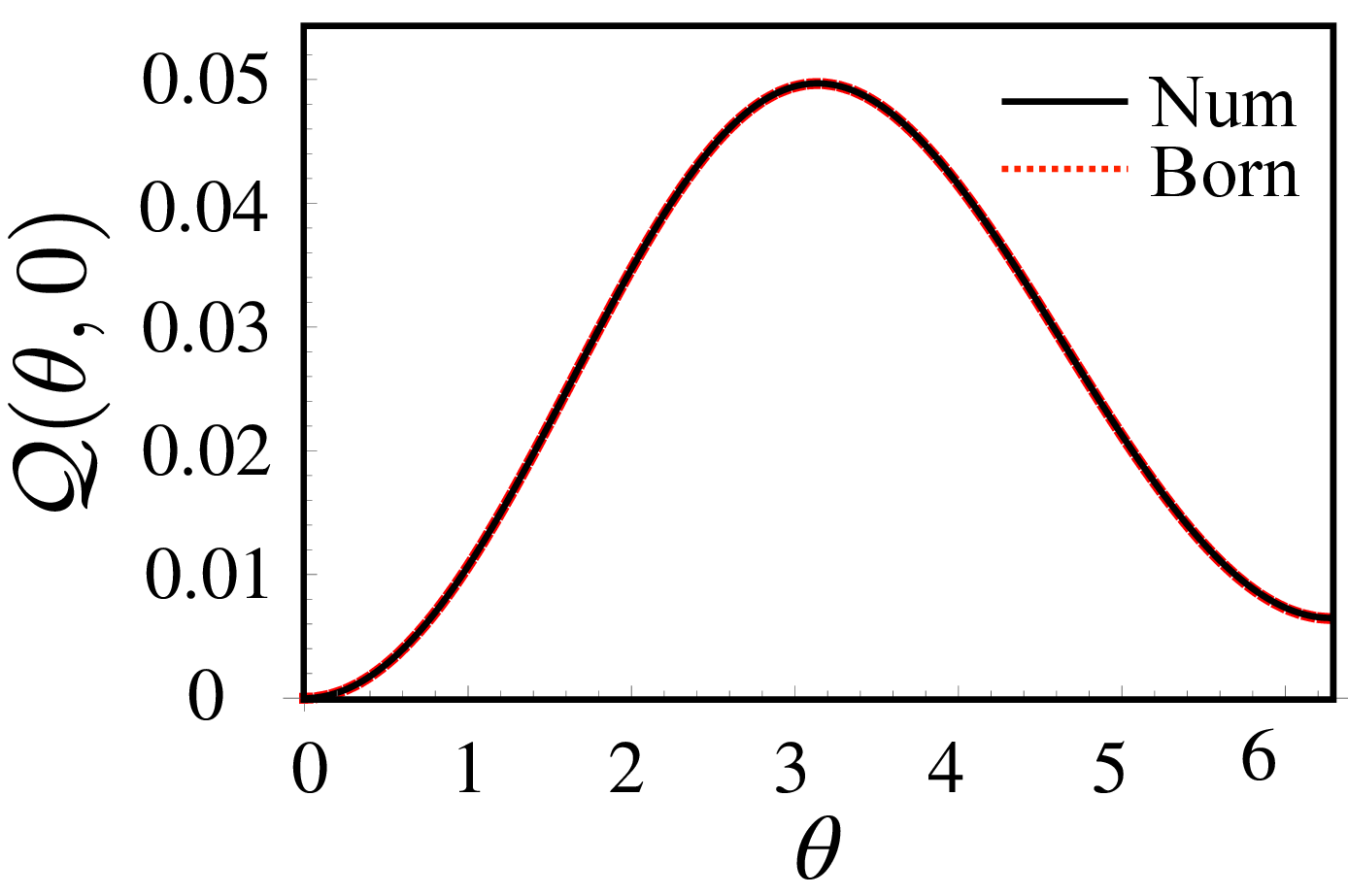}
\caption{
Plot of the heat $\mathcal{Q}(\theta,0)$ versus $\theta$ for $\beta U_0=0.1$ and $r=0.9$. 
}
\label{figheat}
\end{figure}

Third, we also compute the heat flow $\mathcal{Q}(\theta,\delta)$. 
In Fig.\,{\ref{figheat}}, we plot $\mathcal{Q}(\theta,0)$ as a function of $\theta$. 
Since the LHS of Eq.\,\eqref{1st_law_th} is much smaller than $S_{\rm sys}(\theta,\delta)$ and $Q(\theta,\delta)$, $Q(\theta,0)$ is almost the same with  $S_{\rm sys}(\theta,0)$ shown in Fig.\,\,\ref{tildeS}. 
The deviation between the solution of Eq.~\eqref{master} and the result of the Born-like approximation is invisible.

%
\end{document}